\documentclass[11pt,a4paper]{article}
\usepackage{jcappub}

\usepackage{bm}% bold math
\usepackage{mathtools}
\usepackage{mathrsfs}
\usepackage{multirow}
\usepackage{enumerate}
\usepackage{xcolor}
\hypersetup{
    colorlinks=true,
    citecolor=blue,
    linkcolor=violet,
    urlcolor=teal,
}
\newcommand{\rd}{d}
\newcommand{\defi}{\coloneqq}

\newcommand{\pd}{\partial}
\allowdisplaybreaks[4]

\title{Past extendibility and initial singularity in Friedmann--Lema\^{i}tre--Robertson--Walker and Bianchi I spacetimes}

\author{Kimihiro Nomura}%
\emailAdd{knomura@stu.kobe-u.ac.jp}

\author{and Daisuke Yoshida}
\emailAdd{dyoshida@hawk.kobe-u.ac.jp}

\affiliation{%
Department of Physics, Kobe University, Kobe 657-8501, Japan
}%

\preprint{KOBE-COSMO-21-08}

\abstract{We study past-directed extendibility of Friedmann--Lema\^{i}tre--Robertson--Walker (FLRW) and Bianchi type I spacetimes with a scale factor vanishing in the past.
We give criteria for determining whether a boundary for past-directed incomplete geodesics is a parallelly propagated curvature singularity, which cannot necessarily be read off from scalar curvature invariants. 
It is clarified that, for incomplete FLRW spacetime to avoid the singularity, the spacetime necessarily reduces to the Milne universe or flat de Sitter universe toward the boundary. 
For incomplete Bianchi type I spacetime to be free of singularity, it is necessary that the spacetime asymptotically fits into the product of the extendible isotropic geometry (Milne or flat de Sitter) and flat space, or, anisotropic spacetime with specific power law scale factors.
Furthermore, we investigate in detail the time-dependence of the scale factor compatible with the extendibility in both spacetimes beyond the leading order.
}

\arxivnumber{}

\begin{document}

\maketitle
\flushbottom

\section{Introduction}
\label{sec:Introduction}

General Relativity (GR) has been so far the most successful theory describing gravity, as the expected existence of gravitational waves and black holes is supported by observations in recent years \cite{Abbott:2016blz, Akiyama:2019cqa}.
However, its prediction of singularity provides intrinsic evidence that the theory is incomplete anyway.
Indeed, the singularity theorems by Penrose and Hawking state that the occurrence of singularity is unavoidable under some assumptions for energy sources \cite{Penrose:1964wq, Hawking:1969sw, Hawking:1967ju}, so GR would lose the predictability at regions of extremely strong gravity.
With this fact in mind, GR is just an effective theory of gravity applicable at a certain low energy scale.
It is widely expected that the singularity in GR should be appropriately resolved at a more fundamental level, such as by quantum effects of matter and gravity.
Motivated by an idea that some kind of classical gravitation beyond GR incorporating such fundamental effects would effectively hold at some high energy scale, in which the spacetime curvature would be bounded and the singularity could be resolved, there have been efforts to build such theories in the context of cosmology \cite{Mukhanov:1991zn, Brandenberger:1993ef, Moessner:1994jm, Easson:2006jd, Yoshida:2017swb, Quintin:2019orx, Sakakihara:2020rdy} and black holes \cite{Trodden:1993dm, Easson:2002tg, Yoshida:2018kwy}. 
Cosmology with limited curvature is also argued from the perspective of quantum cosmological amplitudes in Ref.\,\cite{Jonas:2021xkx}.
Historically, there have been a lot of studies of spacetime structure without singularity; for example, a regular black hole was first proposed by Bardeen \cite{bardeen1968non}, and subsequent studies can be found, e.g., in Refs.\,\cite{Borde:1996df, Hayward:2005gi,  Bronnikov:2005gm, Bronnikov:2006fu, Ansoldi:2008jw, Bambi:2013ufa}.
Moreover, several attempts have been made to realize Bardeen-like regular black holes as stable solutions in nonlinear electrodynamics with Einstein gravity \cite{AyonBeato:2000zs, Bronnikov:2000vy, Moreno:2002gg, Fan:2016hvf, Nomura:2020tpc}.

To proceed with these approaches, it is crucial to correctly capture when and where the singularity occurs in GR.
For this purpose, let us review the definition of singularity in the literature \cite{Ellis:1977pj, hawking1973large, clarke1973local, Ellis:1974ug, Clarke:1975ph, clarke1982local}.
Suppose that we have a spacetime as a given set of points (not necessarily the entire manifold maximally extended at this moment).
If a causal geodesic reaches the edge of spacetime in a finite affine parameter (a finite proper time for a timelike geodesic), we say that the geodesic is \emph{incomplete} in the spacetime.
We call a set of endpoints of such incomplete geodesics the \emph{boundary} for them.
If at least one component of the Riemann tensor measured in a parallelly propagated (p.p.) tetrad basis along the incomplete geodesics diverges toward the boundary, the spacetime cannot be locally extended beyond it in a way that the component is continuous.
Then we say that the boundary is (\emph{locally}) \emph{inextendible} and there exists a \emph{p.p.~curvature singularity} on the boundary \cite{hawking1973large}.\footnote{More specifically, at least one component of the $r$-th covariant derivative of the Riemann tensor in a p.p. tetrad basis is ill-behaved toward the boundary, the spacetime cannot be locally extended beyond the boundary in a way that the component is continuous.
Then the boundary is said to be a $C^{r}$ \emph{curvature singularity} \cite{Ellis:1977pj}, where we use ``$C^{r}$'' as the differentiability of the Riemann tensor.
In this paper, we intensively argue the $C^{0}$ (continuous) singularity, so we focus only on the behavior of the components of the Riemann tensor (not its derivative at all).
}
(Similar concepts can be established for any causal curve in terms of a generalized affine parameter with respect to a p.p.~tetrad basis.)
On the other hand, if all components of the Riemann tensor in a p.p.~tetrad basis are bounded all the way to the boundary, the spacetime has a local extension beyond it in a way that the components remain well-behaved.
In such a case, we say that the boundary is (\emph{locally}) \emph{extendible}.
We note that the p.p.~curvature singularity can be further classified:
the \emph{scalar} (or \emph{scalar polynomial}; \emph{s.p.}) \emph{curvature singularity} is one that some polynomial of scalar curvature invariants such as the Ricci scalar, Kretschmann scalar, etc., diverges toward the boundary \cite{hawking1973large, Ellis:1977pj}, while the \emph{non-scalar curvature singularity} (or \emph{intermediate singularity}) is one that any scalar curvature invariant is bounded all the way \cite{Ellis:1974ug, Clarke:1975ph, Ellis:1977pj}.
It is also possible that a spacetime cannot have any global extension beyond the boundary although the local geometry is completely well-behaved.
In such a case, the spacetime is said to have a \emph{locally extendible singularity} or \emph{quasi-regular singularity} \cite{Ellis:1974ug, Clarke:1975ph, Ellis:1977pj}, which includes, for example, a conical singularity and the singularity in Taub--NUT spacetime.

In the context of cosmology, the most critical issue is the appearance of initial singularity.
As is well known, in the Friedmann--Lema\^{i}tre--Robertson--Walker (FLRW) universe with a scale factor given by the positive power law of time, $a(t) \propto t^{p}~(p>0)$, any timelike geodesic meets the Big Bang singularity in a finite proper time as going back to the past in general.
On the other hand, as shown by Borde, Guth, and Vilenkin \cite{Borde:2001nh}, in the rapidly expanding universe such as inflation, past-directed causal geodesics are in general incomplete and thus the spacetime has a past boundary.
But, it is not immediately obvious whether the boundary is singular or not.
In Ref.\,\cite{Yoshida:2018ndv}, focusing on the past boundary in the null direction in inflationary cosmology with flat spatial geometry, it is shown that the boundary represents the p.p.~curvature singularity if $\dot{H}/a^{2}$ diverges there, where $a$ is the scale factor and $\dot{H}$ is the time-derivative of the Hubble parameter.
(A similar result can be seen in Refs.\,\cite{FernandezJambrina:2007sx, Fernandez-Jambrina:2016clh} although the possibility of singularity in the approximate flat de Sitter universe is overlooked there.)
This fact implies that whether the inflationary universe has the initial singularity on the past boundary is sensitive to details of the model, but one should note that the analysis in Ref.\,\cite{Yoshida:2018ndv} holds for limited situations, i.e., under assumptions of homogeneity, isotropy, and flat spatial geometry.
Beyond that analysis, it is important to clarify the presence of a past boundary and its (in)extendibility in more general geometry since the behavior of a geodesic congruence is affected by geometrical quantities such as spatial curvature and anisotropy.
The most notable example is the Milne universe, which has an extendible past boundary because of the open spatial curvature (as we will review in Sec.\,\ref{sec:2}).
On the other hand, there has been a belief that, as approaching the initial singularity, spatial anisotropy would exhibit oscillatory behavior and the dynamics at each spatial point might be approximated by homogeneous geometry classified as Bianchi types, which is based on the analysis by Belinsky--Khalatnikov--Lifshitz \cite{Belinsky:1970ew, landau2013classical, Ashtekar:2011ck}. 
By incorporating spatial curvature and anisotropy, we will be able to capture more clearly how the geometry and the existence of singularity are related, which will be useful to identify the singularity of cosmological models and give a guideline for building theories beyond GR to resolve it.

In this paper, we study the local extendibility or p.p.~curvature singularity of a past boundary in spatially homogeneous cosmology, in particular, FLRW spacetime with open, flat, and closed spatial curvature, and Bianchi type I spacetime.
Our main purpose is to give criteria to judge whether the past boundary is locally extendible or inextendible (p.p.~curvature singularity) in terms of the geometrical quantities such as the scale factor and Hubble parameter.
In other words, we will clarify what kind of nontrivial geometry is allowed from a requirement of local extendibility of the past boundary.
We will take the following steps: 
(i) We will investigate whether past-directed comoving and null geodesics reach a set of points where the scale factor (in at least one spatial direction) vanishes at a finite affine parameter.
If so, we can think of spacetime as having a past boundary;
(ii) We will explicitly construct a tetrad basis parallelly propagated along the incomplete geodesic and measure the components of the curvature tensor in that basis. 
Then we will require the finiteness of the components for local extendibility of the boundary.
Conversely, the divergence of the components is equivalent to the p.p.~curvature singularity;
(iii) We will clarify more in detail what the time-dependence of the scale factor or Hubble parameter should be when the past boundary is locally extendible. 

This paper is organized as follows.
In Sec.\,\ref{sec:2}, we review the existence of a past boundary and the extension beyond it in typical examples; the Milne, flat de Sitter, and open de Sitter universes.
The initial p.p.~curvature singularity or local extendibility of a past boundary is investigated for FLRW spacetime with spatial curvature in Sec.\,\ref{sec:Isotropic}, and for Bianchi type I spacetime in Sec.\,\ref{sec:Anisotropic}.
Finally, Sec.\,\ref{sec:Summary} is devoted to the summary and discussion.

\section{Familiar examples of the extendible past boundary}
\label{sec:2}

In this paper, we are interested in the local extendibility of a past boundary in cosmological spacetime, if it exists.
Here, as introduced in Sec.\,\ref{sec:Introduction}, the past boundary stands for a set of points where the scale factor of space vanishes and some past-directed causal geodesic reaches in a finite affine parameter.
The boundary is said to be locally extendible if there is no parallelly propagated (p.p.) curvature singularity, then the apparent pathological behavior of the metric can be removed by introducing some appropriate coordinate system.
In fact, there are some well-known examples of cosmological spacetime with the extendible past boundary.
To capture the idea of this paper, let us review such examples, in particular, the Milne universe, and the open and flat de Sitter universes.

\subsection{The Milne universe}
\label{subsec:Milne}
\begin{figure}[t]
  \centering
  \includegraphics[height=8cm]{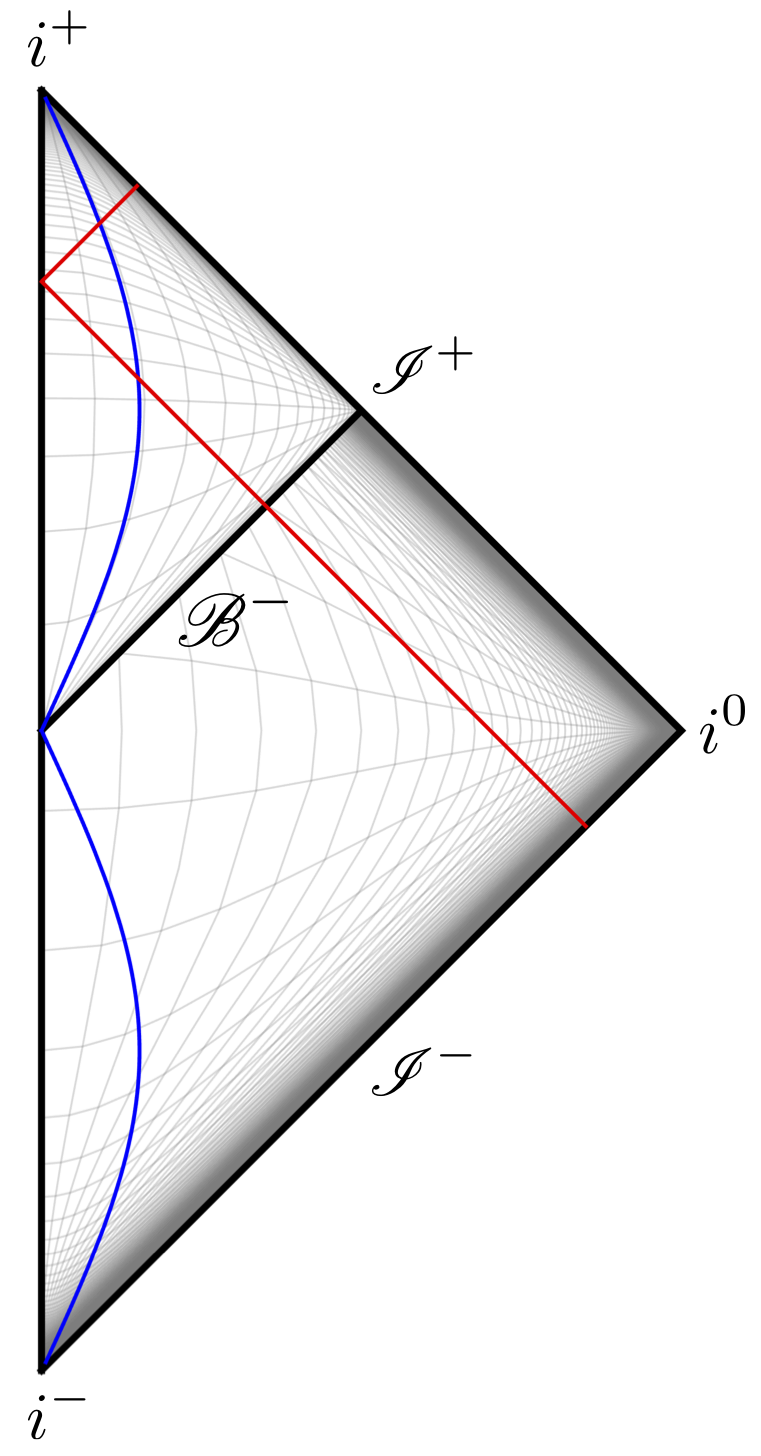}
  \caption{The Penrose diagram of the Milne universe (upper triangle) is a part of the Penrose diagram of Minkowski spacetime (whole triangle). The past-directed comoving (blue) and null (red) curves in the Milne universe reach the past boundary denoted by $\mathscr{B}^{-}$, but they can be extended through it.}
  \label{fig-Milne}
\end{figure}
The Milne universe is known as a solution of Friedmann equations without matter but with negative spatial curvature.
The metric is given by
\begin{align}
ds^{2}_{\text{Milne}} = - \rd t_{+}^{2} + t_{+}^{2} \left( \rd \chi^{2} + \sinh^{2} \chi \, \rd \Omega^{2} \right) \,,
\label{eqK-71}
\end{align} 
where the time coordinate $t_{+}$ is defined within $(0, \infty)$, and $d\Omega^{2} \defi \rd \theta^{2} + \sin^{2} \theta \rd \phi^{2}$ is the metric on the two-sphere.
The metric on the three-hyperboloid appears inside the parentheses.
The whole metric describes the universe expanding linearly in time, i.e., the scale factor is given by $a(t_{+}) = t_{+}$.
The scale factor vanishes at $t_{+} = 0$, and it corresponds to a past boundary where any past-directed causal geodesic reaches in a finite affine parameter.
However, the past boundary is just a coordinate singularity, and the spacetime can be extended beyond it.
Actually, as is well known, the metric \eqref{eqK-71} is just covering inside of the future light cone of some point in Minkowski spacetime.
To see this, we write the Minkowski metric in spherical coordinates,
\begin{align}
ds^{2}_{\text{Minkowski}}
&= - \rd t^{2} + \rd r^{2} + r^{2} \rd \Omega^{2} \,.
\label{eqK-74}
\end{align} 
Then the following coordinate transformation for the Minkowski metric,
\begin{align}
t = t_{+} \cosh \chi \,, \quad
r = t_{+} \sinh \chi \quad (t_{+} > 0)
\label{eqK-75}
\end{align} 
leads to the metric \eqref{eqK-71}.
Since the coordinates introduced in Eq.\,\eqref{eqK-75} satisfy $t > r$, the original metric \eqref{eqK-71} covers only that region of Minkowski spacetime:
\begin{align}
ds^{2}_{\text{Milne}} = ds^{2}_{\text{Minkowski}}|_{t>r} \,.
\label{eqK-76} 
\end{align} 
As shown in Fig.\,\ref{fig-Milne}, the Penrose diagram of the Milne universe can be drawn as that part of Minkowski spacetime.

\subsection{The flat and open de Sitter universes}
\label{subsec:deSitter}

\begin{figure}[t]
  \centering
  \includegraphics[height=5cm]{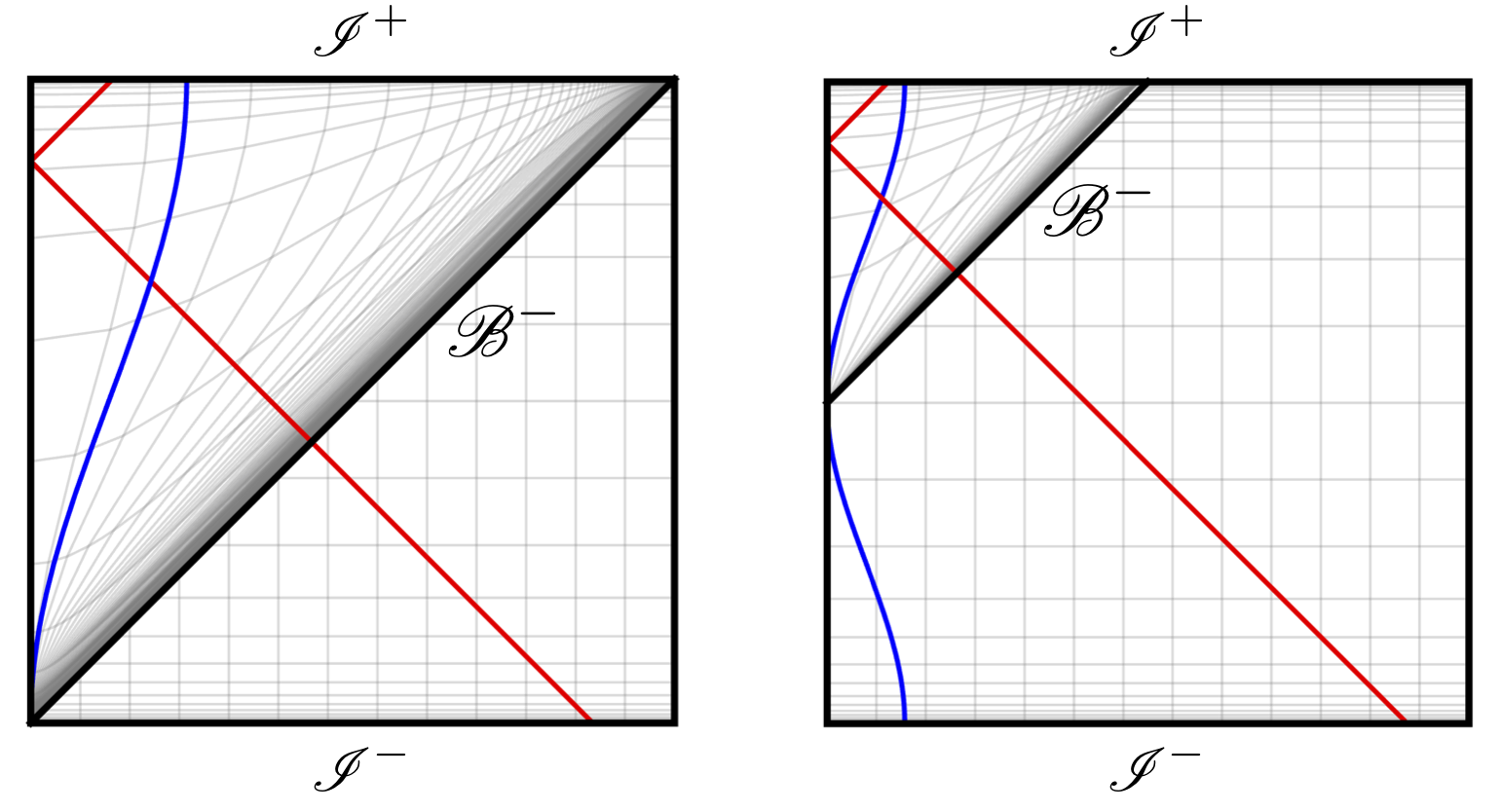}
  \caption{Left: the Penrose diagram of the flat de Sitter universe is given by the triangle covering the upper left half. The comoving (blue) curve is complete. The null (red) curve reaches the past boundary $\mathscr{B}^{-}$ of the flat de Sitter universe in a finite affine parameter, but it can be extended through it.
  Right: the Penrose diagram of the open de Sitter universe is given by the upper left triangle. The comoving (blue) and null (red) curves reach the past boundary $\mathscr{B}^{-}$ of the open de Sitter universe at a finite affine parameter, but they can be extended through it.
  In both figures, the entire square corresponds to the closed de Sitter universe.}
  \label{fig-dS}
\end{figure}

As is well known, de Sitter spacetime, which is the maximally symmetric spacetime with a positive cosmological constant, can be represented in various coordinate charts, such as spatially open, flat, and closed charts.
In the closed chart, the metric of de Sitter spacetime reads
\begin{align}
ds^{2}_{\text{closed-dS}}
&= - \rd t^{2} + \frac{\cosh^{2}(H_{\Lambda} t) }{H_{\Lambda}^{2}} \left( \rd \chi^{2} + \sin^{2} \chi \, \rd \Omega^{2} \right) \,.
\label{eqdS-1}
\end{align}
Here, a constant $H_{\Lambda}$ is related with the cosmological constant $\Lambda$ by $H_{\Lambda} = \sqrt{\Lambda / 3}$, and the time coordinate $t$ is defined from $- \infty$ to $+ \infty$.
The closed de Sitter universe is geodesically complete, i.e., any causal geodesic in the universe has no end at some finite affine parameter. 
On the other hand, de Sitter spacetime also can be expressed as the spatially flat universe as
\begin{align}
ds^{2}_{\text{flat-dS}}
&= - \rd t_{1}^{2} + \frac{e^{2H_{\Lambda}t_{1}}}{H_{\Lambda}^{2} }
\left( \rd x_{1}^{2} + \rd x_{2}^{2} + \rd x_{3}^{2} \right) \,.
\label{eqdS-4}
\end{align}
The flat de Sitter universe has a past boundary in the null direction since there exist past-directed incomplete null geodesics which reach the points $t_{1} = - \infty$, where the scale factor $e^{H_{\Lambda} t_{1}}/H_\Lambda$ vanishes, at finite values of the affine parameter.
We can also write the de Sitter metric as the spatially open universe,
\begin{align}
ds^{2}_{\text{open-dS}}
&= - \rd t_{2}^{2} + \frac{\sinh^{2}(H_{\Lambda}t_{2})}{H_{\Lambda}^{2}} \left( \rd \xi_{1}^{2} + \sinh^{2} \xi_{1}^{2} \rd \Omega^{2} \right) \,,
\label{eqdS-6}
\end{align} 
where $\rd \Omega^{2} = \rd \xi_{2}^{2} + \sin^{2} \xi_{2} \rd \xi_{3}^{2}$.
In this metric, the scale factor $\sinh(H_{\Lambda}t_{2})/H_{\Lambda}$ vanishes at $t_{2} = 0$, and there is a past boundary for causal geodesics.
The essential point here is that the past boundaries in the flat and open de Sitter universes do not represent any singularity since the appearance of the boundaries is just due to introducing the incomplete coordinate charts.
The flat and open de Sitter universes can be regarded as parts of the closed one, and thus there is no obstacle to extend them beyond the past boundaries.
This should be clear from the Penrose diagrams shown in Fig.\,\ref{fig-dS}.

\section{Past extendibility and initial singularity in FLRW spacetime}
\label{sec:Isotropic}

\subsection{Setup in FLRW spacetime}
\label{subsec:Iso-Setup}

In this section, we consider spatially homogeneous and isotropic spacetime and analyze the presence or absence of initial singularity.
In Ref.\,\cite{Yoshida:2018ndv}, the case of spatially flat spacetime is studied, while in this paper we allow a spacetime to have a constant spatial curvature.
Such a spacetime can be described by the Friedmann--Lema\^{i}tre--Robertson--Walker (FLRW) metric with coordinates $x^{\mu} = (t, \chi , \theta, \phi)$ as
\begin{align}
\rd s^{2} = 
g_{\mu\nu} \rd x^{\mu} \rd x^{\nu} = - \rd t^2 + a^{2}(t) \left( \rd \chi^2 + \Phi_{k}^{2}(\chi) \rd \Omega^2 \right) \,.
\label{eqK-1}
\end{align}
Here, $a(t)$ is the scale factor, $\rd \Omega^{2} \defi \rd \theta^{2} + \sin^{2} \theta \rd \phi^{2}$ is the metric on the unit two-sphere, and $\Phi_{k}(\chi)$ is defined by
\begin{align}
 \Phi_{k}(\chi) \defi 
\begin{cases}
 \sin \chi & (k = 1) \,,\\
 \chi &(k = 0) \,, \\
 \sinh \chi &(k = -1) \,,
\end{cases}
\label{eqK-2}
\end{align}
where $k$ is the normalized spatial curvature.
We take the time coordinate $t$ to be defined within a domain $(t_{i}, t_{f})$ as will be mentioned below.
The coordinate $\chi$ is defined with $\chi \in (0, \infty)$ for $k = 0$ and $k = -1$, while $\chi \in (0, \pi)$ for $k = +1$.

Throughout this section, we are interested in the initial singularity in an expanding universe.
We therefore assume that the scale factor $a(t)$ vanishes at the initial time $t = t_{i}$ as going back to the past without loss of generality: 
\begin{align}
  \lim_{t \to t_{i}}a(t) = 0\,.
  \label{eqK-2.2}
\end{align}
Our aim is to study whether this pathological behavior of the metric at $t_{i}$ leads to a singularity or not.
Note that we do not assume that $t_{i}$ is finite, so it can be negative infinity. 
In this regard, our analysis includes singularities in a wide class not covered in Ref.\,\cite{FernandezJambrina:2006hj} where singularities at a finite comoving time in FLRW cosmology are focused on.
Since we are concerned only about the early behavior, we take the end of the time domain $t_{f}$ to be finite arbitrarily so that the scale factor takes some positive and finite value toward that time: $\lim_{t \rightarrow t_{f}} a(t) = \text{finite}$.
A similar analysis for the singularity in a contracting universe can be performed simply by flipping the time coordinate, $t \to -t$.

As mentioned in Sec.\,\ref{sec:Introduction}, if the spacetime has incomplete causal geodesics which reach the points of $a=0$ (or equivalently, $t=t_{i}$) at finite values of the affine parameter, a set of the points amounts to the past boundary for the geodesics.
However, it is not immediately obvious whether the boundary represents a parallelly propagated (p.p.) curvature singularity (i.e., the geodesics cannot be extended beyond it anymore) or not (i.e., the geodesics can be extended beyond it at least locally).
Below, we will study the property of the boundary in the past-directed comoving and null directions, separately.

\subsection{Past comoving direction in FLRW spacetime}
\label{subsec:Iso-timelike}

Let us consider a comoving timelike curve represented in the coordinates $x^{\mu} = (t, \chi, \theta, \phi)$ as
\begin{align}
 x^{\mu}(\tau) = (\tau + t_0 , \chi_{0}, \theta_{0}, \phi_{0}) \,, \label{eqK-21}
\end{align}
with constants $t_{0}, \chi_{0}, \theta_{0}$, and $\phi_{0}$.
Here, $\tau$ is used as a parameter of the curve.
The tangent vector of this curve is given by
\begin{align}
 \boldsymbol{u} = \frac{d x^{\mu}(\tau)}{ d \tau} \boldsymbol{\partial}_{\mu} = \boldsymbol{\partial}_{t} \,,
\label{eqK-22}
\end{align}
where $\bm{\pd}_{\mu}$ is the coordinate basis with respect to $x^{\mu}$.
The tangent vector $\bm{u}$ satisfies the geodesic equation,
\begin{align}
\nabla_{\boldsymbol{u}} \boldsymbol{u}  
= 0 \,,
\label{eqK-23}
\end{align}
where $\nabla_{\bm{u}}$ represents the covariant derivative along the vector $\bm{u}$.
Therefore, the comoving curve \eqref{eqK-21} is a geodesic and $\tau$ is its affine parameter.
The geodesic reaches $a = 0$ at $\tau = t_{i} - t_{0}$.
If $t_{i} = -\infty$, since the affine parameter can run to $- \infty$ until it reaches $a = 0$, the spacetime is past complete for the comoving geodesic.
On the other hand, if $t_{i}$ is finite, the comoving geodesic reaches $a=0$ at a finite affine parameter and thus there is a past boundary for the comoving geodesic.
Then, the metric \eqref{eqK-1} is past incomplete for the comoving geodesic.

In the past comoving incomplete case, i.e., $t_{i} = \text{finite}$, we are concerned about whether the geodesic can be extended beyond the boundary, at least locally.
As mentioned in Sec.\,\ref{sec:Introduction}, a useful way to judge the local extendibility is to check the components of the Riemann tensor in a parallelly propagated (p.p.) tetrad basis along the geodesic.
One can see that the following one-forms
\begin{subequations} 
\begin{align}
 \boldsymbol{e}^{0} &= \rd t \label{eqK-24}\,,\\
 \boldsymbol{e}^{1} &= a(t) \rd \chi \,,
 \label{eqK-25}\\
 \boldsymbol{e}^{2} &= a(t) \Phi_{k}(\chi) \rd \theta \,,
 \label{eqK-26}\\
 \boldsymbol{e}^{3} &= a(t) \Phi_{k}(\chi) \sin \theta \rd \phi \,,
 \label{eqK-27}
 \end{align}
\end{subequations}
are parallel along the comoving geodesic;
\begin{align}
\nabla_{\boldsymbol{u}} \boldsymbol{e}^{M} = 0 \,,
\label{eqK-28}
\end{align}
where $M = 0,1,2,3$. 
Therefore, the tetrad basis $\{ \bm{e}^{M} \}$ serves as the p.p.~tetrad basis along the comoving geodesic. 
If at least one component of the Riemann tensor in this basis diverges in the limit to $t_{i}$, it is indicated that there exists a p.p.~curvature singularity on the boundary and thus we cannot extend the spacetime beyond it.
Otherwise, the spacetime is extendible along the comoving geodesic, at least locally.
 
Since the FLRW metric is conformally flat, the independent components of the Riemann tensor are described only by the Ricci tensor.
In the p.p.~tetrad basis, the components of the Ricci tensor can be evaluated as
\begin{align}
 R_{MN} \bm{e}^{M} \otimes \bm{e}^{N} 
 &= - 2 \left(  \dot{H}  - \frac{k}{a^2} \right) \boldsymbol{e}^{0} \otimes \boldsymbol{e}^{0} 
 \notag \\
 &\quad +
 \left( \left( \dot{H} - \frac{k}{a^2} \right) + 3 \left(H^2 + \frac{k}{a^2} \right) \right) \eta_{MN} \boldsymbol{e}^{M} \otimes \boldsymbol{e}^{N}\,,
\label{eqK-29}
\end{align}
where a dot stands for the derivative with respect to the coordinate time $t$, $H \defi \dot{a}/a$ is the Hubble parameter, and $\eta_{MN} = \text{diag} (-1,+1,+1,+1)$.
Thus the incomplete comoving geodesic is at least locally extendible if
\begin{align}
&\lim_{t \rightarrow t_{i}} 
\left( \dot{H} - \frac{k}{a^2}   \right) = \text{finite} \,,\label{eqK-30} \\
&\lim_{t \rightarrow t_{i}}   \left( H^2 + \frac{k}{a^2}   \right) =  \text{finite} \,.
\label{eqK-31}
\end{align}
These conditions might be well known because they are equivalent to require the finiteness of scalar curvature invariants.
In fact, the Ricci scalar $R$ and the Ricci tensor squared ${R}_{\mu\nu} {R}^{\mu\nu}$ are calculated as
\begin{align}
R &= 6 \left(\dot{H} - \frac{k}{a^2}\right) + 12 \left( H^2 + \frac{k}{a^2}\right) \,,
\label{eqK-32} \\
R_{\mu\nu}R^{\mu\nu}
&= 12 \left(\dot{H} - \frac{k}{a^2}\right)^{2} + 36 \left(\dot{H} - \frac{k}{a^2}\right) \left( H^2 + \frac{k}{a^2}\right) + 36 \left( H^2 + \frac{k}{a^2}\right)^{2} \,.
\label{eqK-33}
\end{align}
Comparing these with Eqs.\,\eqref{eqK-30} and \eqref{eqK-31}, it turns out that if FLRW spacetime is not locally extendible along a comoving geodesic, it is always accompanied by a scalar curvature singularity.
Therefore, one does not need to care about the possibility of a non-scalar curvature singularity.

Given a perfect fluid with the energy density $\rho$ and pressure $p$, from Friedmann equations we have
\begin{align}
\dot{H} - \frac{k}{a^{2}} &= - 4 \pi G (\rho + p) \,,
\label{eqK-34}\\
H^{2} + \frac{k}{a^{2}} &= \frac{8 \pi G}{3} \rho \,,
\label{eqK-35}
\end{align} 
where $G$ is the gravitational constant.
From the conditions \eqref{eqK-30} and \eqref{eqK-31}, to make the past boundary for comoving geodesics locally extendible, the fluid must satisfy
\begin{align}
\lim_{t \to t_{i}} \rho = \text{finite} \,, \quad
\lim_{t \to t_{i}} p = \text{finite}\,.
\label{eqK-36}
\end{align}

\subsection{Past null direction in FLRW spacetime}
\label{subsec:Iso-null}

Let us turn to consider a null curve written in the coordinates $x^{\mu} = (t, \chi, \theta, \phi)$ as 
\begin{align}
 x^{\mu}(\lambda) = \left( t(\lambda) + t_{0}, - \int d\lambda \,\frac{\pd_{\lambda} t(\lambda)}{a} + \chi_{0}, \theta_{0}, \phi_{0}\right)
\label{eqK-41}
\end{align}
with a parameter $\lambda$ and constants $t_{0}$, $\chi_{0}$, $\theta_{0}$, and $\phi_{0}$.
This null curve is actually a null geodesic because its tangent vector,
\begin{align}
 \boldsymbol{k} = 
 \frac{\pd x^{\mu}(\lambda)}{\pd \lambda} \bm{\pd}_{\mu} =
 \pd_{\lambda} t \, \boldsymbol{\partial}_{t} - \frac{\pd_{\lambda} t}{a} \, \boldsymbol{\partial}_{\chi} \,,
 \label{eqK-42}
\end{align}
satisfies the geodesic equation
\begin{align}
 \nabla_{\boldsymbol{k}} \boldsymbol{k} = \left(\frac{\partial_{\lambda} \partial_{\lambda}t}{\partial_{\lambda} t} + \frac{\partial_{\lambda} a}{a}\right) \boldsymbol{k}\,. 
\label{eqK-43}
\end{align}
The relation between $t$ and $\lambda$ should be determined by requiring that $\lambda$ is an affine parameter, i.e., the right-hand side of Eq.\,\eqref{eqK-43} vanishes.
This requirement is satisfied by
\begin{align}
\rd \lambda = a \, \rd t \,.
\label{eqK-45}
\end{align}
Then the tangent vector is given by
\begin{align}
\bm{k} = \frac{1}{a} \bm{\pd}_{t} - \frac{1}{a^{2}} \bm{\pd}_{\chi} \,.
\label{eqK-45-2}
\end{align}
The past-directed completeness of the null geodesic can be checked by evaluating the difference of the affine parameter between the initial time $t_{i}$ and some finite time $t_{r}$:  
\begin{align}
 \lambda(t_{i}) - \lambda(t_{r}) = \int^{t_{i}}_{t_{r}} dt \, a(t) =
\begin{cases}
 - \infty\text{: complete} ,\\
 \text{finite: incomplete} .
\end{cases}
\label{eqK-46}
\end{align}
Note that the null geodesic is always incomplete when a comoving geodesic is incomplete, namely when $t_{i}$ is finite.
On the other hand, it is possible that the spacetime is complete in the past comoving direction but incomplete in the past null direction, i.e., a null geodesic reaches the past boundary at a finite affine parameter.

We focus on a case that the spacetime is incomplete in the past null direction.
To judge the local extendibility beyond the past boundary, we need a tetrad basis parallelly propagated along the null geodesic.
The simplest tetrad $\{ \boldsymbol{e}^{M} \}$ defined by Eqs.\,\eqref{eqK-24}--\eqref{eqK-27} does not satisfy this property because $\boldsymbol{e}^{0}$ and $\boldsymbol{e}^{1}$ are not parallel along the null geodesic,
\begin{align}
\nabla_{\boldsymbol{k}} \boldsymbol{e}^{0} & = H \rd \chi \,,
\label{eqK-47}\\
\nabla_{\boldsymbol{k}} \boldsymbol{e}^{1} & = \frac{H}{a} \rd t \,,
\label{eqK-48}
\end{align}
while $\boldsymbol{e}^{2}$ and $\boldsymbol{e}^{3}$ are parallel: $\nabla_{\boldsymbol{k}} \boldsymbol{e}^{2}  =   \nabla_{\boldsymbol{k}} \boldsymbol{e}^{3} = 0$.
Now, let us introduce a set of one-forms $\{ \bm{l}, \bm{n}, \bm{e}^{2}, \bm{e}^{3} \}$, where $\bm{l}$ and $\bm{n}$ are null one-forms defined by
\begin{subequations} 
\begin{align}
\bm{l} &= \frac{1}{\sqrt{2} \, a}(\bm{e}^{0} + \bm{e}^{1})
\,, \label{eqK-48-1}\\
\bm{n} &= \frac{a}{\sqrt{2}}(\bm{e}^{0} - \bm{e}^{1}) \,.
\label{eqK-48-2}
\end{align} 
\end{subequations} 
Then, the covariant derivatives of $\bm{l}$ and $\bm{n}$ along $\bm{k}$ vanish:
\begin{align}
\nabla_{\bm{k}} \bm{l} &= 0 \,,
\label{eqK-48-3}\\
\nabla_{\bm{k}} \bm{n} &= 0 \,.
\label{eqK-48-4}
\end{align} 
Therefore, we can use the set $\{ \bm{l}, \bm{n}, \bm{e}^{2}, \bm{e}^{3} \}$ as the p.p.~tetrad basis along the null geodesic.
We can argue the presence or absence of the p.p.~curvature singularity on the past boundary by checking the components of the Ricci tensor in this tetrad basis.
In terms of the tetrad basis $\{ \bm{l}, \bm{n}, \bm{e}^{2}, \bm{e}^{3} \}$, the Ricci tensor \eqref{eqK-29} is expressed as
\begin{align}
&R_{MN} \bm{e}^{M} \otimes \bm{e}^{N}
\notag \\
&= - \left( \dot{H} - \frac{k}{a^{2}} \right) \left(  a^{2} \bm{l} \otimes \bm{l} + \frac{1}{a^{2}} \bm{n} \otimes \bm{n}  + \bm{l} \otimes \bm{n} + \bm{n} \otimes \bm{l} \right)
\notag \\
&\quad+ \left( \left( \dot{H} - \frac{k}{a^2} \right) + 3 \left(H^2 + \frac{k}{a^2} \right) \right) \left(\bm{e}^{2} \otimes \bm{e}^{2} + \bm{e}^{3} \otimes \bm{e}^{3} -\bm{l} \otimes \bm{n} - \bm{n} \otimes \bm{l} \right) \,.
\label{eqK-48-5}
\end{align} 
Focusing on the coefficient of $\bm{n} \otimes \bm{n}$, we can see that the finiteness of the Ricci tensor requires a condition
\begin{align}
 \lim_{t \rightarrow  t_{i}} \frac{1}{a^2} \left( \dot{H} - \frac{k}{a^2}\right) = \text{finite} \,.
\label{eqK-55}
\end{align}
This condition is stronger than Eq.\,\eqref{eqK-30} in a sense that the scale factor squared, which approaches zero as $t \to t_{i}$, appears in the denominator additionally. 
This result, which corresponds to a generalization of the result in Ref.\,\cite{Yoshida:2018ndv} to general spatial curvature, is consistent with Refs.\,\cite{FernandezJambrina:2006hj, FernandezJambrina:2007sx}.
In terms of the energy density $\rho$ and pressure $p$ of the perfect fluid fulfilling the spacetime, Eq.\,\eqref{eqK-55} requires 
\begin{align}
\lim_{t \to t_{i}} \frac{\rho + p}{a^{2}} = \text{finite}\,.
\label{eqK-56}
\end{align} 
Now defining the equation-of-state parameter $w$ by $p = w\rho$, and taking account of $\rho \propto a^{-3(1+w)}$ from the energy conservation of the fluid, Eq.\,\eqref{eqK-56} can be translated to
\begin{align}
w = -1 \quad \text{or} \quad w \leq - \frac{5}{3} \quad \text{as} ~ t\to t_{i}\,.
\label{eqK-57}
\end{align} 

The results obtained so far for FLRW spacetime are summarized in Table \ref{table1} and as follows.
When past-directed incomplete causal geodesics exist, the fulfillment of the conditions \eqref{eqK-31} and \eqref{eqK-55} ensures that the curvature tensor in the p.p.~tetrad basis for the incomplete geodesics is well-behaved and thus the past boundary $(a=0)$ is at least locally extendible.
Otherwise, the past boundary is a p.p.~curvature singularity.
If any of the conditions \eqref{eqK-30} and \eqref{eqK-31} is violated, there is a scalar curvature singularity on the past boundary.
If both of the conditions \eqref{eqK-30} and \eqref{eqK-31} are satisfied but the condition \eqref{eqK-55} is not, there is a non-scalar curvature singularity on the past boundary in the null direction, which is not detectable by focusing only on scalar curvature invariants.

\begin{table}[tb]
  \centering
\begin{tabular}{c|c|c|c|l}
$t_i$ & $\lambda(t_{i})$        & $\dot{H}-\frac{k}{a^2}, H^2+\frac{k}{a^2}$ & $\frac{1}{a^2}\left(\dot{H} - \frac{k}{a^2}\right)$ & \multicolumn{1}{c}{singularity} \\ \hline
\multirow{4}{*}{\begin{tabular}[c]{@{}c@{}}$- \infty$\\ \end{tabular}} 
      & $-\infty$               & --                       & --                      & complete \\ \cline{2-5} 
      & \multirow{3}{*}{finite} & \multirow{2}{*}{finite}  & finite                  & coordinate \\ \cline{4-5} 
      &                         &                          & $\pm \infty$            & non-scalar \\ \cline{3-5} 
      &                         & $\pm \infty$             & --                      & scalar \\ \hline
\multirow{3}{*}{\begin{tabular}[c]{@{}c@{}}finite\\  \end{tabular}}    
      & \multirow{3}{*}{finite} & \multirow{2}{*}{finite}  & finite                  & coordinate \\ \cline{4-5}
      &                         &                          & $\pm \infty$            & non-scalar \\ \cline{3-5} 
      &                         & $\pm \infty$             & --                      &scalar                      
\end{tabular}
\caption{Classification of the past boundary corresponding to $a = 0 $ in FLRW spacetime:
$t_i$ is the initial time when the scale factor vanishes, and $\lambda(t_i)$ is the initial affine parameter along a null geodesic.
In the rightmost row, ``complete'' means that both comoving and null geodesics are past complete and there is no past boundary, ``coordinate'' means that the past boundary is a coordinate singularity and it is extendible at least locally, ``non-scalar'' means that the boundary is a non-scalar curvature singularity, and ``scalar'' means that the boundary is a scalar curvature singularity. }
\label{table1}
\end{table}

\subsection{Extendible past boundary in FLRW spacetime}
\label{subsec:Iso-3}

So far, we have investigated FLRW spacetime in which past-directed causal geodesics reach the boundary ($a=0$) at a finite affine parameter.
We have obtained the conditions for the past boundary to be at least locally extendible as
\begin{align}
\lim_{t \to t_{i}} \left( H^{2} + \frac{k}{a^{2}} \right) &= \text{finite}
\label{eqK-81} \,,\\
\lim_{t \to t_{i}} \frac{1}{a^{2}} \left( \dot{H} - \frac{k}{a^{2}} \right) &= \text{finite} \,.
\label{eqK-82}
\end{align} 
In this subsection, let us study more in detail when these extendibility conditions are satisfied.
In particular, we will identify the sign of the spatial curvature and the time-dependence of the Hubble parameter and scale factor in the limit toward the boundary so that the spacetime is at least locally extendible beyond it.
For this purpose, first note that combining the above two conditions gives the following differential equation for the Hubble parameter,
\begin{align}
\dot{H} + H^{2} = Q
\quad (t \to t_{i})
\label{eqK-83}
\end{align} 
with some finite constant $Q$.
In Appendix \ref{appA}, we solve the equation \eqref{eqK-83} and classify the asymptotic behavior of the solutions.
Below, we will utilize the results obtained there to find the time-dependence of the Hubble parameter and scale factor in the extendible spacetime.

As already mentioned, there are two types for past incomplete spacetime; (i) incomplete in both comoving and null directions, i.e., $t_{i} = \text{finite}$, (ii) complete in the comoving direction but incomplete in the null direction, i.e., $t_{i} = - \infty$ and $\lambda(t_{i}) = \text{finite}$.
We consider these cases separately.
We will see that the extendible cases are summarized as Table \ref{tab:extendibleFLRW}.

\subsubsection{Comoving and null incomplete case: $t_{i} = \text{finite}$}
\label{Iso-3-tfin}

In a case of $t_{i} = \text{finite}$, the scale factor $a(t)$ vanishes at a finite time $t_{i}$ as going back to the past.
For simplicity, we take $t_{i} = 0$ without loss of generality.
Note that then the Hubble parameter $H(t)$ must diverge to positive infinity in the limit $t \to 0$.
This fact can be easily understood by integrating the definition of the Hubble parameter, $H \defi \dot{a}/a$, from $t=0$ to $t_{f}$;
\begin{align}
\int_{0}^{t_{f}} dt \, H(t) = \ln a(t_{f}) - \ln a(0) \,.
\label{eqK-84}
\end{align} 
The right-hand side diverges to positive infinity since $a(0) = 0$ from our assumption.
The interval of integration in the left-hand side is finite, and thus $H(t)$ must diverge to positive infinity in the limit $t \to 0$.
Then in order to satisfy the condition \eqref{eqK-81}, we need
\begin{align}
k = -1 \,.
\label{eqK-85}
\end{align}
That is, the past comoving and null incomplete FLRW spacetime has to be spatially open to avoid the singularity on the boundary.

To identify the time-dependence of the Hubble parameter explicitly in the extendible case, we need to find solutions to Eq.\,\eqref{eqK-83} which diverge to positive infinity in the limit $t \to 0$.
The solutions are found in Appendix \ref{appA}, and as summarized in Eq.\,\eqref{eqA-7}, their leading behavior around $t=0$ must be
\begin{align}
H(t) = \frac{1}{t} + \mathcal{O}(t^0) \,.
\label{eqK-86.2}
\end{align} 
Then we have the leading behavior of the scale factor around $t=0$ as $a(t) = t+ \mathcal{O}(t^2)$.
Now let us constrain the subleading behavior.
We expand the scale factor around $t=0$ as
\begin{align}
a(t) = t + \sum_{p=2}^{\infty} \frac{a^{(p)}}{p!} t^{p} 
\label{eqK-87-2}
\end{align} 
where $a^{(p)}$ are finite coefficients independent of $t$.
With $k = -1$, then we have 
\begin{align}
H^{2} + \frac{k}{a^{2}} &= \frac{2a^{(2)}}{t} + \mathcal{O}(t^{0}) \,,
\label{eqK-87-3}\\
\frac{1}{a^{2}}\left( \dot{H} - \frac{k}{a^{2}} \right) &= - \frac{a^{(2)}}{t^{3}} + \frac{3(a^{(2)})^{2}}{2t^{2}} + \frac{1}{t} \left( \frac{a^{(4)}}{6} + \frac{a^{(2)}a^{(3)}}{3} - \frac{3(a^{(2)})^{3}}{2} \right) + \mathcal{O}(t^{0}) \,.
\label{eqK-87-4}
\end{align} 
Thus the extendibility conditions \eqref{eqK-81} and \eqref{eqK-82} constrain the time-dependence of the scale factor as
\begin{align}
a^{(2)} = a^{(4)} = 0 \,,
\label{eqK-87-5}
\end{align} 
or equivalently,
\begin{align}
a(t) &= t + \frac{a^{(3)}}{3!} t^{3} + \mathcal{O}(t^{5}) \,.
\label{eqK-87-6}
\end{align}
Then the time-dependence of the Hubble parameter is determined as
\begin{align}
H(t) &= \frac{1}{t} + \frac{a^{(3)}}{3}t + \mathcal{O}(t^3) \,.
\label{eqK-87-7}
\end{align}  

In summary, if FLRW spacetime has a past boundary in both comoving and null directions at $t=0$, i.e., the scale factor vanishes at $t=0$, the spacetime must be spatially open ($k=-1$) and the scale factor and Hubble parameter must obey Eqs.\,\eqref{eqK-87-6} and \eqref{eqK-87-7} to make the past boundary extendible at least locally.
Note that the extendibility requires that the metric asymptotically takes the form of the Milne universe,
\begin{align}
  ds^{2} 
  \simeq -\rd t^{2} + t^{2} (\rd \chi^{2} + \sinh^{2} \chi \rd \Omega^{2}) \quad (t \to 0)
  \label{eqK-89}
  \end{align} 
at the leading order, and it is also sensitive to the subleading behavior of the scale factor with respect to time.
Of course, the open de Sitter universe \eqref{eqdS-6} is included in the extendible case since the scale factor satisfies the above condition.

\subsubsection{Comoving complete but null incomplete case: $t_{i} = - \infty$ and $\lambda(t_{i}) = \text{finite}$}
\label{subsubsec:Iso-3-tinf}

In a case of $t_{i} = - \infty$ while $\lambda(t_{i}) = \text{finite}$, where $\lambda(t_i)$ is the affine parameter of a null geodesic at $t_i=-\infty$ (see Eq.\,\eqref{eqK-46}), we need to find $H$ satisfying Eq.\,\eqref{eqK-83} in the limit $t \to - \infty$ to make the boundary extendible.
As mentioned in the end of Appendix \ref{appA}, such $H$ should converge to some constant in the limit.
Thus we need
\begin{align}
\lim_{t \to -\infty} H(t) = \bar{H}
\label{eqK-91}
\end{align} 
with some non-negative constant $\bar{H}$. 
Then, the conditions \eqref{eqK-81} and \eqref{eqK-82} are satisfied only if
\begin{align}
k=0 \,.
\label{eqK-92}
\end{align}
That is, the spacetime should be spatially flat to avoid the singularity.

However, it is possible that even if $k=0$ and the Hubble parameter converges to a constant as $t \to - \infty$, the extendibility conditions are violated due to the subleading behavior of the Hubble parameter (or scale factor) with respect to $t$.
This is the case, for instance, in the Starobinsky inflation model \cite{Yoshida:2018ndv}.
Below, let us generically constrain the subleading behavior from the requirement of the extendibility.
We consider some situations such that the Hubble parameter deviates from the constant $\bar{H}$ as
\begin{align}
H(t) = \bar{H} + f(t) \,,
\label{eqK-94}
\end{align} 
where we assume that $f(t)$ vanishes as $t \to -\infty$ so that the condition \eqref{eqK-91} still holds.
Then it is obvious that the condition \eqref{eqK-81} is satisfied with $k =0$.
But, $\dot{H}(t) = \dot{f}(t)$ is nontrivial, so the condition \eqref{eqK-82} might be violated.
Below, as examples, let us consider the cases such that the deviation of the Hubble parameter from the constant, $f(t) = H(t) - \bar{H}$, is approximately given by the negative power of $-t$, and the exponential of $-t$.

\paragraph{Deviation by $(-t)^{-p}$.}
Let us consider the Hubble parameter deviated from some constant $\bar{H}$ by the negative power of $-t$ in the past enough:
\begin{align}
H(t) \simeq \bar{H} + \frac{b}{(-t)^{p}} \quad (p>0)\,,
\label{eqK-101}
\end{align} 
where $b$ is some non-zero constant.
First, we assume $\bar{H} > 0$, then there is always the past boundary in the null direction (see Refs.\,\cite{Borde:2001nh,Yoshida:2018ndv}).
Our concern is whether the boundary is locally extendible or not.
Then the scale factor can be obtained as
\begin{align}
a(t) \simeq \begin{dcases}
  e^{\bar{H} t} \left(1 + \frac{b}{p-1} \frac{1}{(-t)^{p-1}} \right)
  &(p > 1)\,,\\
  e^{\bar{H}t} (-t)^{-b}
  &(p = 1)\,,\\
  e^{\bar{H} t}\exp\left( \frac{-b}{1-p} (-t)^{1-p} \right)
  &(0 < p < 1)\,,
\end{dcases} 
\label{eqK-102}
\end{align} 
where irrelevant terms as $t \to -\infty$ are ignored.
Then the conditions to avoid the scalar curvature singularity, Eqs.\,\eqref{eqK-30} and \eqref{eqK-31}, are satisfied with $k = 0$.
But, the dominant time dependence of $\dot{H}/a^2$ as $t \to -\infty$ is approximately given by
\begin{align}
\frac{\dot{H}}{a^{2}}
\sim bp \,e^{-2\bar{H}t} 
\qquad (t \to -\infty)\,,
\label{eqK-103}
\end{align} 
where some factor with the minor time dependence such as the power of $t$ is suppressed.
This quantity necessarily diverges as $t \to - \infty$ because of the factor of $e^{-2\bar{H}t}$, which comes from $a^{2}$ in the denominator.
Thus the condition \eqref{eqK-82} is violated, i.e., the non-scalar curvature singularity occurs on the past boundary.

Next, let us turn to set $\bar{H} = 0$ in Eq.\,\eqref{eqK-101}.
Then we need $b > 0$ to obtain an expanding universe, and the scale factor is approximately given by
\begin{align}
  a(t) \simeq \begin{dcases}
    1 + \frac{b}{p-1} \frac{1}{(-t)^{p-1}}
    &(p > 1)\,,\\
    (-t)^{-b}
    &(p = 1)\,,\\
    \exp\left( \frac{-b}{1-p} (-t)^{1-p} \right)
    &(0 < p < 1)\,,
  \end{dcases} 
  \label{eqK-104}
\end{align} 
where irrelevant terms as $t \to -\infty$ are ignored again.
When $p>1$, the scale factor does not vanish as $t \to -\infty$, which is out of our interest.
For the case of $p = 1$, as studied in Ref.\,\cite{FernandezJambrina:2007sx}, the completeness in the past null direction can be checked through Eq.\,\eqref{eqK-46} as
\begin{align}
\lambda(-\infty) - \lambda(t_{r})
&\simeq \int_{t_{r}}^{-\infty} dt \,(-t)^{-b}
= \begin{cases}
- \infty\text{: complete} &(0 < b \leq 1) \,,
\\
\text{finite: incomplete} &(1<b)\,.
\end{cases} 
\label{eqK-105}
\end{align} 
Thus the case of $0 < b \leq 1$ with $p=1$ has no past boundary, which is also out of our interest.
In the past null incomplete case, i.e. $b>1$ with $p=1$, we can see that the quantity
\begin{align}
\frac{\dot{H}}{a^{2}} \simeq b(-t)^{2b-2}
\label{eqK-106}
\end{align} 
diverges as $t \to - \infty$ and thus the past boundary is the non-scalar curvature singularity.
Finally, for the case of $0 < p < 1$, the scale factor vanishes as $t \to -\infty$.
To ensure that there is the past boundary in the null direction, note that we can take some reference time $t_{r} (<0)$ so that the relation
\begin{align}
a(t) \simeq \exp\left( \frac{-b}{1-p} (-t)^{1-p} \right) < \frac{1}{t^{2}}
\label{eqK-107}
\end{align} 
always holds within $t \in (-\infty, t_{r})$.
Then, integrating both sides leads to
\begin{align}
\lambda(t_{r}) - \lambda(-\infty)
\simeq \int_{-\infty}^{t_{r}} dt \, a(t) 
< \int_{-\infty}^{t_{r}} \frac{1}{t^{2}} \, dt
= \frac{1}{|t_{r}|}\,.
\label{eqK-108}
\end{align}  
Therefore, a past-directed null geodesic has a finite affine length, i.e., there exists the past boundary in the null direction.
The boundary represents the non-scalar curvature singularity since
\begin{align}
\frac{\dot{H}}{a^{2}} \simeq \frac{bp}{(-t)^{p+1}} \exp\left( \frac{2b}{1-p} (-t)^{1-p} \right)
\label{eqK-109}
\end{align} 
diverges as $t \to -\infty$.

We can conclude that the deviation of the Hubble parameter from a constant by the negative power of $-t$ as Eq.\,\eqref{eqK-101} keeps the scalar curvature invariants finite when $k=0$, but leads to the non-scalar curvature singularity on the past boundary, if it exists, i.e., the spacetime is past null incomplete.

\paragraph{Deviation by $e^{ct}~(c>0)$.}
As another example of the deviation like Eq.\,\eqref{eqK-94}, let us consider the Hubble parameter given by
\begin{align}
H(t) \simeq \bar{H} + d\, e^{ct} \quad (c>0)
\label{eqK-110}
\end{align} 
in the past enough, where $c(>0)$ and $d$ are constants.
Here we take $\bar{H}$ to be positive, then there exists the past boundary for null geodesics.\footnote{If one sets $\bar{H} = 0$ in Eq.\,\eqref{eqK-110} to consider the Hubble parameter given by $H(t) = d e^{ct}$, the direct integration gives $a(t) = e^{(d/c)e^{ct}}$. This scale factor does not vanish as $t \to -\infty$, which is out of our interest.}
Corresponding to the above Hubble parameter, the scale factor is approximately given by
\begin{align}
a(t) \simeq e^{\bar{H}t} + \frac{d}{c} e^{(\bar{H}+c)t}\,,
\label{eqK-111}
\end{align} 
where irrelevant terms as $t \to -\infty$ are ignored.
With $k=0$, the scalar curvature singularity does not occur on the boundary since the conditions \eqref{eqK-30} and \eqref{eqK-31} are satisfied.
On the other hand, to avoid the non-scalar curvature singularity, the quantity
\begin{align}
\frac{\dot{H}}{a^{2}} \sim cd\, e^{-(2\bar{H} - c)t} 
\label{eqK-112}
\end{align} 
must be finite as $t \to -\infty$, i.e., 
\begin{align}
c \geq 2\bar{H}\,.
\label{eqK-113}
\end{align} 
The key reason why the singularity is avoided under Eq.\,\eqref{eqK-113} is that $\dot{H}$ exponentially approaches zero in the limit $t \to -\infty$ as fast as or faster than $a^{2}$ and thus $\dot{H}/a^{2}$ does not diverge.
It is easily understood that the deviation of the Hubble parameter from a constant by a function for which $\dot{H}$ vanishes faster than $a^{2} \sim e^{2\bar{H}t}$ as $t \to -\infty$, e.g. $H(t) = \bar{H} + e^{-\alpha t^{2}}$ with some positive constant $\alpha$, does not violate the conditions \eqref{eqK-81} and \eqref{eqK-82}, and thus the past boundary remains at least locally extendible when $k=0$.

From the above analysis, it follows that when the past boundary in the null direction is extendible, the metric must asymptotically take the following form, 
\begin{align}
ds^{2}
\simeq - \rd t^{2} + e^{2\bar{H}t} (\rd \chi^{2} + \chi^{2} \rd \Omega^{2}) \quad (t \to - \infty)
\label{eqK-93}
\end{align} 
at the leading order, which is the metric of the flat de Sitter universe.

% \usepackage{multirow}
\begin{table}[t]
  \centering
  \begin{tabular}{l|l|l}
  \multicolumn{3}{l}{\textbf{Extendible FLRW}}                                                                                                                        \\ \hline
                                            & spatial curvature             & scale factor                          \\ \hline
  comoving and null incomplete:             & \multirow{2}{*}{open: $k=-1$} & \multirow{2}{*}{$a(t)=t + \alpha t^3 + \mathcal{O}(t^5)$} \\ 
  $a(0) = 0$                              &                               &                                        \\ \hline
  null incomplete: $a(-\infty) =0$        & \multirow{2}{*}{flat: $k=0$}  & $a(t) = e^{\bar{H}t} + \alpha e^{(\bar{H}+c)t}+ \cdots$  \\ 
  and $\int^{-\infty}a(t)\,dt = \text{finite}$ &                               & with $c \geq 2\bar{H}$ \\ \hline
  \end{tabular}
  \caption{FLRW spacetimes with at least locally extendible past boundary (an apparent coordinate singularity) at $a=0$. Here, $\alpha$ is an arbitrary constant.}
  \label{tab:extendibleFLRW}
  \end{table}

\section{Past extendibility and initial singularity in Bianchi I spacetime}
\label{sec:Anisotropic}

In the previous section, we considered the spatially homogeneous and isotropic universe. 
As a next step, it is natural to ask how the extendibility conditions for a past boundary change if the assumption of isotropy is removed.
In this section, we extend the work in the previous section to the spatially homogeneous but anisotropic universe.
In particular, we consider Bianchi type I spacetime where each spatial direction has a different scale factor in general.

\subsection{Setup in Bianchi I spacetime}

Bianchi type I spacetime is known as the simplest case of spatially homogeneous spacetime.
Assuming that the metric is diagonalized, it is given by
\begin{align}
ds^{2}  
= - \rd t^{2} + a_{x}^{2}(t) \rd x^{2} + a_{y}^{2}(t) \rd y^{2} + a_{z}^{2}(t) \rd z^{2} \,.
\label{eqI-1}
\end{align} 
Here, $a_{j}(t)~(j=x,y,z)$ is the scale factor in each spatial direction.
We use $H_{j} \defi \dot{a}_{j}/a_{j}$ as the Hubble parameter in each direction, where a dot stands for the derivative with respect to the coordinate time $t$.

Since we are interested in the presence or absence of initial singularity, we consider a situation that the scale factor in at least one spatial direction vanishes as going back to the past.
With a time domain $(t_{i},t_{f})$, without loss of generality, we can assume 
\begin{align}
\lim_{t \to t_{i}} a_{x}(t) = 0 \,.
\label{eqI-2}
\end{align} 
Equivalently, it can be thought that we \emph{define} the initial time $t_{i}$ as the time when one of the scale factors, $a_{x}$, vanishes irrespective of whether $a_{y}$ and $a_{z}$ vanish at $t_{i}$ or not.
Notice that we do not assume that $t_{i}$ is finite, i.e., $t_{i}$ can be negative infinity.
Since we are interested only in the initial singularity, we take $t_{f}$ as some finite time so that all scale factors are finite in the limit to $t_{f}$.

Note that reversing the time coordinate in our arguments, $t \to -t$, corresponds to the analysis for singularity in the future. 
In Ref.\,\cite{Cataldo:2015eja}, the future singularity at a finite comoving time in Bianchi type I spacetime is studied. 
There, the singularity where at least one of the scale factors diverges is intensively considered, but the situations such that another scale factor vanishes, which we deal with in this paper, are also included. 
In this regard, our analysis has an overlap with that one.
But, we consider a wide class of singularity in a sense that our analysis also includes the situations such that at least one of the scale factors approaches zero at \emph{infinite} coordinate time.

\subsection{Past comoving direction in Bianchi I spacetime}
\label{Anisotropic:timelike}

First, let us consider a comoving timelike curve expressed in the coordinates $x^{\mu} = (t, x, y, z)$ as
\begin{align}
x^{\mu}(\tau) = (\tau + t_{0}, x_{0}, y_{0}, z_{0}) \,,
\label{eqI-01}
\end{align} 
where $t_{0}$, $x_{0}$, $y_{0}$, and $z_{0}$ are some constants, and $\tau$ is used as a parameter of the curve.
The tangent vector of the curve is given by
\begin{align}
\bm{u} = \frac{dx^{\mu}(\tau)}{d \tau} \bm{\pd}_{\mu} = \bm{\pd}_{t} \,.
\label{eqI-02}
\end{align} 
Then, the covariant derivative of the tangent vector along itself vanishes, $\nabla_{\bm{u}} \bm{u} =0$.
Thus, the curve \eqref{eqI-01} is a comoving geodesic and $\tau$ is its affine parameter.
Since the scale factor in the $x$-direction, $a_{x}$, goes to zero in the limit to $\tau = t_{i} - t_{0}$ by the assumption, the comoving geodesic \eqref{eqI-01} is past complete if $t_{i} = -\infty$.
On the other hand, if $t_{i}$ is finite, the spacetime is incomplete in the past comoving direction and $a_{x} = 0$ represents the past boundary.

We focus on a case that the comoving geodesic reaches $a_{x}=0$ at finite $t_{i}$ and thus the spacetime is incomplete in the past comoving direction.
To discuss the local extendibility beyond the past boundary ($a_{x}=0$), we look at the components of the Riemann tensor in a tetrad basis parallelly propagated along the geodesic.
For this purpose, we set up a simple tetrad as
\begin{subequations} 
\begin{align}
{\bm{e}}^{0} &= \rd t \,,
\label{eqI-03.1} \\
{\bm{e}}^{1} &= a_{x}(t) \rd x \,,
\label{eqI-03.2} \\
{\bm{e}}^{2} &= a_{y}(t) \rd y \,,
\label{eqI-03.3} \\
{\bm{e}}^{3} &= a_{z}(t) \rd z \,.
\label{eqI-03.4}
\end{align} 
\end{subequations} 
This simple tetrad is parallelly propagated along the comoving geodesic \eqref{eqI-01}, i.e., $\nabla_{\bm{u}} {\bm{e}}^{M} =0~(M=0,1,2,3)$.
Thus we can discuss the local extendibility in the past comoving direction by using this parallelly propagated (p.p.) tetrad basis.

For convenience, we separate the components of the Riemann tensor into the Ricci and Weyl tensors.
In the p.p.~tetrad basis, the Ricci tensor reads
\begin{align}
  R_{MN} \bm{e}^{M} \otimes \bm{e}^{N}
  &= R_{00} {\bm{e}}^{0}\otimes {\bm{e}}^{0} + R_{11} {\bm{e}}^{1} \otimes {\bm{e}}^{1} + R_{22} {\bm{e}}^{2} \otimes {\bm{e}}^{2} + R_{33} {\bm{e}}^{3} \otimes {\bm{e}}^{3} \,,
  \label{eqI-04}
\end{align}
where
\begin{align}
R_{00} &= - \dot{H}_{x} - H_{x}^{2} - \dot{H}_{y} - H_{y}^{2} - \dot{H}_{z} - H_{z}^{2} \,,
\label{eqI-04-2} \\
R_{11} &= H_{x}^{2} + H_{x} H_{y} + H_{z}H_{x} + \dot{H}_{x} \,,
\label{eqI-04-3} \\
R_{22} &= H_{x}H_{y} + H_{y}^{2} + H_{y}H_{z} + \dot{H}_{y} \,,
\label{eqI-04-4} \\
R_{33} &= H_{z}H_{x} + H_{y}H_{z} + H_{z}^{2} + \dot{H}_{z} \,,
\label{eqI-04-5} 
\end{align}
while the Weyl tensor reads
\begin{align}
&C_{KLMN} \bm{e}^{K} \otimes \bm{e}^{L} \otimes \bm{e}^{M} \otimes \bm{e}^{N} 
\notag \\ 
&= C_{0101}  \big( ({\bm{e}}^{0} \otimes {\bm{e}}^{1} \otimes {\bm{e}}^{0} \otimes {\bm{e}}^{1} + \text{3 perms})
 - ( {\bm{e}}^{2} \otimes {\bm{e}}^{3} \otimes {\bm{e}}^{2} \otimes {\bm{e}}^{3} + \text{3 perms} ) \big)
\notag \\
&\quad + \text{(cyclic permutations of } (1,2,3))\,,
\label{eqI-05}
\end{align}  
where ``perms'' represents the permutations of bases with appropriate sign allowed by symmetry and antisymmetry of the tensor, and
\begin{align}
C_{0101} &= -\frac{1}{6} ( 2\dot{H}_{x} - \dot{H}_{y} - \dot{H}_{z} + 2H_{x}^{2} - H_{y}^{2} - H_{z}^{2} 
 + 2H_{y}H_{z} - H_{z}H_{x} - H_{x}H_{y} ) \,.
\label{eqI-05-1}
\end{align} 
If the components of the Ricci and Weyl tensors listed above are not bounded in the limit to $t_{i}$, the past boundary is singular and thus we cannot extend the spacetime beyond it.
In fact, we can simplify the extendibility conditions by combining some components of the Ricci and Weyl tensors.
For example, some calculations give
\begin{align}
\frac{R_{00}}{6} - \frac{R_{11}}{6} + \frac{R_{22}}{3} + \frac{R_{33}}{3} - C_{0101}
&= H_{y}H_{z} \,,
\label{eqI-05-2}\\
- \frac{R_{00}}{3} + \frac{R_{11}}{3} - \frac{R_{22}}{6} - \frac{R_{33}}{6} - C_{0101} &= \dot{H}_{x} + H_{x}^{2} \,.
\label{eqI-05-3}
\end{align}
Therefore, the finiteness of each component of the Ricci and Weyl tensors in the limit to $t_{i}$ implies
\begin{align}
\lim_{t \to t_{i}} (\dot{H}_{j} + H_{j}^{2}) &= \text{finite} \,,
\label{eqI-05-4}\\
\lim_{t \to t_{i}} (H_{j} H_{k}) &= \text{finite} \quad (j \neq k) \,.
\label{eqI-05-5}
\end{align}
Inversely, if the conditions \eqref{eqI-05-4} and \eqref{eqI-05-5} are satisfied, we can see that all components of the Ricci and Weyl tensors are finite.
Thus it is equivalent that the comoving incomplete spacetime can be extended beyond the past boundary and that the conditions \eqref{eqI-05-4} and \eqref{eqI-05-5} are satisfied.
On the other hand, if $\dot{H}_{j} + H_{j}^{2}$ and/or $H_{j} H_{k}$ diverge as $t \to t_{i}$, there is the p.p.~curvature singularity in the past comoving direction.
Note that we have four components of the Ricci tensor in Eq.\,\eqref{eqI-04} and three components of the Weyl tensor in Eq.\,\eqref{eqI-05} with one constraint $C_{0101} + C_{0202} + C_{0303} = 0$. 
Thus, the number of the independent components, six, coincides with the number of the conditions given by Eqs.\,\eqref{eqI-05-4} and \eqref{eqI-05-5}.

It is instructive to compare the above results with the regularity condition of scalar curvature invariants.
In this spacetime, for example, we have
\begin{align}
R &= 2 (\dot{H}_{x} + H_{x}^{2} + \dot{H}_{y} + H_{y}^{2} + \dot{H}_{z} + H_{z}^{2} + H_{x}H_{y} + H_{y}H_{z} + H_{z}H_{x}) \,,
\label{eqI-06}\\
R^{\mu\nu}R_{\mu\nu}
&= (\dot{H}_{x} + H_{x}^{2} + \dot{H}_{y} + H_{y}^{2} + \dot{H}_{z} + H_{z}^{2} )^2 + (\dot{H}_x + H_x^2 + H_x H_y + H_z H_x)^2 
\notag \\
&\quad + (H_x H_y + \dot{H}_y + H_y^2 + H_y H_z)^2 + (H_z H_x + H_y H_z + \dot{H}_z +  H_z^2 )^2 \,,
\label{eqI-07-2}\\
R^{\mu\nu\rho\sigma} R_{\mu\nu\rho\sigma}
&= 4 \big( (\dot{H}_{x}+H_{x}^{2})^{2} + (\dot{H}_{y}+H_{y}^{2})^{2} + (\dot{H}_{z}+H_{z}^{2})^{2} + H_{x}^{2}H_{y}^{2} + H_{y}^{2}H_{z}^{2} + H_{z}^{2}H_{x}^{2} \big) \,.
\label{eqI-07-3}
\end{align}
We can see that the divergence of the components of the Riemann tensor in the p.p.~tetrad basis along the comoving geodesic always induces the divergence of the scalar curvature invariants.
Thus the p.p.~curvature singularity in the comoving direction necessarily implies the presence of the scalar curvature singularity. 
In other words, there never exists a non-scalar curvature singularity such that any scalar curvature invariant is bounded but at least one component of the Riemann tensor in the p.p.~basis diverges along the comoving direction.

\subsection{Past null direction in Bianchi I spacetime}
\label{Anisotropic:null}

Let us turn to discuss the presence or absence of a past boundary and local extendibility beyond it in the past null direction.
We consider a past-directed null curve along the $x$-direction represented in the coordinates $x^{\mu} = (t, x, y, z)$  as 
\begin{align}
x^{\mu}(\lambda)
= \left( t(\lambda) + t_{0}, 
- \int \rd \lambda \, \frac{\pd_{\lambda}t(\lambda)}{a_{x}} + x_{0} , y_{0}, z_{
0}
\right) \,.
\label{eqI-5}
\end{align}
Here, $\lambda$ is used as a parameter of the curve.
In fact, we can see that the tangent vector of the curve,
\begin{align}
\bm{k} = \frac{dx^{\mu}(\lambda)}{d\lambda } \bm{\pd}_{\mu}
= \pd_{\lambda}t \,\bm{\pd}_{t} - \frac{\pd_{\lambda}t}{a_{x} } \bm{\pd}_{x} \,,
\label{eqI-5-2}
\end{align} 
is null, and satisfies the geodesic equation,
\begin{align}
\nabla_{\bm{k}} \bm{k}
= \left( \frac{\pd_{\lambda} \pd_{\lambda} t}{\pd_{\lambda} t } + \frac{\pd_{\lambda} a_{x}}{a_{x} }\right) \bm{k} \,.
\label{eqI-6}
\end{align} 
If the right-hand side of the geodesic equation vanishes, we can think of $\lambda$ as an affine parameter.
It is fulfilled by
\begin{align}
\rd \lambda = a_{x} \, \rd t \,.
\label{eqI-10}
\end{align} 
Given this parametrization, the tangent vector is written as
\begin{align}
\bm{k} = \frac{1}{a_{x}} \bm{\pd}_{t} - \frac{1}{a_{x}^{2}} \bm{\pd}_{x} \,.
\label{eqI-11}
\end{align} 
The past-directed completeness of the null geodesic can be checked from whether the affine parameter can run to $- \infty$ until it reaches $a_{x} = 0$ ($t = t_{i}$). 
From Eq.\,\eqref{eqI-10}, we have
\begin{align}
\lambda(t_{i}) - \lambda(t_{r}) = \int_{t_{r}}^{t_{i}} dt \, a_{x}(t)
= \begin{cases}
- \infty\text{: complete} ,\\
\text{finite: incomplete} ,
\end{cases} 
\label{eqI-10-2}
\end{align} 
where $t_{r}$ is some finite time.
If $\lambda(t_{i}) = - \infty$, the spacetime is past complete along the null geodesic. 
From the assumption $\lim_{t \to t_{i}} a_{x} = 0$, it requires $t_{i} = -\infty$, then the spacetime is also past complete in the comoving direction.
Thus $\lambda(t_{i}) = - \infty$ means that the spacetime is geodesically complete and has no boundary in the past, which is out of our interest.
On the other hand, if the affine parameter $\lambda$ takes some finite value in the limit to $t_{i}$, the spacetime is incomplete in the past null direction and there is a past boundary for the null geodesic.
Notice that when the spacetime is past null incomplete, there are two possibilities; the spacetime is comoving complete, i.e. $t_{i} = - \infty$, or the spacetime is comoving incomplete, i.e. $t_{i} = \text{finite}$.

We focus on a case of $ \lambda(t_{i}) = \text{finite}$, in which there is a past boundary for the null geodesic.
Then we need to check the components of the Riemann tensor in a p.p.~tetrad basis along the null geodesic to discuss the local extendibility of the boundary.
For this purpose, by using the tetrad $\{\bm{e}^{M}\}$ in Eqs.\,\eqref{eqI-03.1}--\eqref{eqI-03.4}, we set up two null one-forms as
\begin{subequations} 
\begin{align}
\bm{l} &= \frac{1}{\sqrt{2} \, a_{x}}(\bm{e}^{0} + \bm{e}^{1})
\,, \label{eqI-11.1}\\
\bm{n} &= \frac{a_{x}}{\sqrt{2}}(\bm{e}^{0} - \bm{e}^{1}) \,.
\label{eqI-11.2}
\end{align} 
\end{subequations} 
One can see that the covariant derivatives of one-forms $\{ \bm{l}, \bm{n}, \bm{e}^{2}, \bm{e}^{3} \}$ along the tangent vector $\bm{k}$ all vanish.
Thus we can use this set of one-forms as a p.p.~tetrad basis along the null geodesic \eqref{eqI-5}.
In this basis, the Ricci and Weyl tensors read 
\begin{align}
R_{MN} {\bm{e}}^{M} \otimes {\bm{e}}^{N}
&= \frac{R_{00}+R_{11}}{2} \left( a_{x}^{2} \bm{l} \otimes \bm{l} + \frac{1}{a_{x}^{2}} \bm{n} \otimes \bm{n} \right)
 + \frac{R_{00} - R_{11}}{2} (\bm{l} \otimes \bm{n} + \bm{n} \otimes \bm{l} )
\notag \\
&\quad + R_{22} {\bm{e}}^{2} \otimes {\bm{e}}^{2} + R_{33} {\bm{e}}^{3} \otimes {\bm{e}}^{3} \,,
\label{eqI-13}
\end{align} 
and
\begin{align}
&C_{KLMN} {\bm{e}}^{K} \otimes {\bm{e}}^{L} \otimes {\bm{e}}^{M} \otimes {\bm{e}}^{N} 
\notag \\ 
&= C_{0101} \bigg( (\bm{l} \otimes \bm{n} \otimes \bm{l} \otimes \bm{n} + \text{3 perms})
 - ({\bm{e}}^{2} \otimes {\bm{e}}^{3} \otimes {\bm{e}}^{2} \otimes {\bm{e}}^{3} + \text{3 perms})
\notag \\
&\qquad\qquad - \frac{1}{2} (\bm{l} \otimes {\bm{e}}^{2} \otimes \bm{n} \otimes {\bm{e}}^{2} + \text{7 perms})
 - \frac{1}{2} (\bm{l} \otimes {\bm{e}}^{3} \otimes \bm{n} \otimes {\bm{e}}^{3} + \text{7 perms}) \bigg)
\notag \\
&\quad + \frac{C_{0202} - C_{0303}}{2} \bigg( a_{x}^{2} (\bm{l} \otimes {\bm{e}}^{2} \otimes \bm{l} \otimes {\bm{e}}^{2} + \text{3 perms})
- a_{x}^{2} (\bm{l} \otimes {\bm{e}}^{3} \otimes \bm{l} \otimes {\bm{e}}^{3} + \text{3 perms}) 
\notag \\
&\qquad\qquad + 
\frac{1}{a_x^2} (\bm{n} \otimes {\bm{e}}^{2} \otimes \bm{n} \otimes {\bm{e}}^{2} + \text{3 perms})
 - \frac{1}{a_x^2} ( \bm{n} \otimes {\bm{e}}^{3} \otimes \bm{n} \otimes {\bm{e}}^{3} + \text{3 perms}) \bigg)
\,,
\label{eqI-14}
\end{align}  
respectively.
Here, $R_{00}, R_{11} , \dots, C_{0101}, \dots$ can be read off from Eqs.\,\eqref{eqI-04-2}--\eqref{eqI-04-5} and \eqref{eqI-05-1}. 

Let us collect conditions for the past boundary to be locally extendible.
First, by imposing the finiteness of the components of the Ricci and Weyl tensors in the limit to $t_{i}$ (where $a_{x} \to 0$), we have the same conditions as Eqs.\,\eqref{eqI-05-4} and \eqref{eqI-05-5} as
\begin{align}
\lim_{t \to t_{i}} (\dot{H}_{j} + H_{j}^{2}) &= \text{finite} \,,
\label{eqI-17}\\
\lim_{t \to t_{i}} (H_{j} H_{k}) &= \text{finite} \quad (j \neq k) \,,
\label{eqI-18}
\end{align}
but note that here $t_{i}$ can be $- \infty$.
In addition, if the $\bm{n} \otimes \bm{n}$ component of the Ricci tensor and the $\bm{n} \otimes \bm{e}^{2} \otimes \bm{n} \otimes \bm{e}^{2}$ (or $\bm{n} \otimes \bm{e}^{3} \otimes \bm{n} \otimes \bm{e}^{3}$) component of the Weyl tensor are bounded as $t \to t_{i}$, we have
\begin{align}
\lim_{t \to t_{i}} \left( - \frac{R_{00} + R_{11}}{2 a_{x}^{2}} - \frac{C_{0202} - C_{0303}}{a_x^2} \right) &= 
\lim_{t \to t_{i}} \frac{1}{a_{x}^{2}}(\dot{H}_{y} + H_{y}^{2} - H_{x}H_{y}) 
= \text{finite} \,,
\label{eqI-19} \\
\lim_{t \to t_{i}} \left( - \frac{R_{00} + R_{11}}{2 a_{x}^{2}} + \frac{C_{0202} - C_{0303}}{a_x^2} \right) &= 
\lim_{t \to t_{i}} \frac{1}{a_{x}^{2}}(\dot{H}_{z} + H_{z}^{2} - H_{x}H_{z}) 
= \text{finite} \,.
\label{eqI-20} 
\end{align} 
Inversely, given the Hubble parameters obeying Eqs.\,\eqref{eqI-17}--\eqref{eqI-20}, we can see that all components of the Ricci and Weyl tensors are well-behaved in the limit to $t_{i}$.
Thus the conditions \eqref{eqI-17}--\eqref{eqI-20} serve as necessary and sufficient conditions for the past boundary to be locally extendible.
If at least one of the conditions \eqref{eqI-17}--\eqref{eqI-20} is not satisfied, there exists a p.p.~curvature singularity.
As already mentioned, if any of the conditions \eqref{eqI-17} and \eqref{eqI-18} is violated, there is a scalar curvature singularity.
If both of the conditions \eqref{eqI-17} and \eqref{eqI-18} are satisfied but any of the conditions \eqref{eqI-19} and \eqref{eqI-20} is violated, there is a non-scalar curvature singularity.
These results are summarized in Table \ref{table2}.
Note that the conditions \eqref{eqI-19} and \eqref{eqI-20} cannot be obtained by focusing only on the scalar curvature invariants such as Eqs.\,\eqref{eqI-06}--\eqref{eqI-07-3}.
They would be severer conditions than Eqs.\,\eqref{eqI-17} and \eqref{eqI-18} in a sense that $a_{x}$, which approaches zero as $t \to t_{i}$, appears in the denominators.

\begin{table}[tb]
  \centering
\begin{tabular}{c|c|c|c|l}
$t_i$	&$\lambda(t_{i})$	&$\dot{H}_j + H_j^2$, $H_j H_k$&	$\frac{1}{a_{x}^{2}}(\dot{H}_{y,z} + H_{y,z}^{2} - H_{x}H_{y,z})$	&singularity \\ \hline
\multirow{4}{*}{$- \infty$}	&$-\infty$	&--	&--	& complete	\\ \cline{2-5} 
&\multirow{3}{*}{finite}		&\multirow{2}{*}{finite}	& finite	& coordinate	\\ \cline{4-5} 
&	&	&$\pm \infty$	&non-scalar	\\ \cline{3-5} 
&	&$\pm \infty$	& --	& scalar		\\ \hline
\multirow{3}{*}{finite}	&\multirow{3}{*}{finite}	&\multirow{2}{*}{finite}	&finite	&coordinate	\\ \cline{4-5}
&	&	&$\pm \infty$	&non-scalar	\\ \cline{3-5} 
&	&$\pm \infty$	&--	&scalar                      
\end{tabular}
\caption{Classification of the past boundary corresponding to $a_{x} = 0 $ in Bianchi type I spacetime:
$t_i$ is the initial time when the scale factor $a_x$ vanishes, and $\lambda(t_i)$ is the initial affine parameter of a null geodesic in the $x$-direction.
In the rightmost row, ``complete'' means that both comoving and null geodesics are past complete and there is no past boundary, ``coordinate'' means that the past boundary is a coordinate singularity and it is extendible at least locally, ``non-scalar'' means that the boundary is a non-scalar curvature singularity, and ``scalar'' means that the boundary is a scalar curvature singularity. }
\label{table2}
\end{table}

\subsection{Extendible past boundary in Bianchi I spacetime}
\label{subsec:Aniso-4}

In the above analysis, we obtained conditions for the past boundary to be at least locally extendible as
\begin{align}
\lim_{t \to t_{i}} (\dot{H}_{j} + H_{j}^{2}) &= \text{finite} \,,
\label{eqI-21}\\
\lim_{t \to t_{i}} (H_{j} H_{k}) &= \text{finite} \quad (j \neq k) \,,
\label{eqI-22}\\
\lim_{t \to t_{i}} \frac{1}{a_{x}^{2}}(\dot{H}_{y,z} + H_{y,z}^{2} - H_{x}H_{y,z}) 
&= \text{finite} \,.
\label{eqI-23} 
\end{align} 
Here, $t_{i}$ is the time when $a_{x}$ vanishes.
Based on these conditions, let us study more in detail how the Hubble parameters and scale factors should behave in the limit $t \to t_{i}$ to avoid the p.p.~curvature singularity.
First of all, let us focus on the condition \eqref{eqI-21}. 
It implies that in order to make a past boundary extendible, it is important to solve the following differential equation,
\begin{align}
\dot{H}_{j} + H_{j}^{2} = Q_{j}
\quad \text{as }t \to t_{i}
\label{eqI-24}
\end{align} 
with a constant $Q_{j}$.
In Appendix \ref{appA}, we find solutions of this equation and classify their flows with time. 
In this section, we will utilize the results to identify the time dependence of the Hubble parameters and scale factors.

Again, there are two types for past incomplete spacetime; (i) incomplete in both comoving and null directions, i.e. $t_{i} = \text{finite}$, (ii) complete in the comoving direction but incomplete in the null direction, i.e. $t_{i} = - \infty$ but $\lambda(t_{i}) = \text{finite}$.
We consider these cases separately.
We will see that the extendible past boundary occurs in the cases listed in Table \ref{tab:extendibleBianchiI}.

\subsubsection{Comoving and null incomplete case: $t_{i} = \text{finite}$}

Here we consider a case that the scale factor in the $x$-direction, $a_{x}$, vanishes at a finite time $t_{i}$.
For simplicity, we take $t_{i} = 0$ without loss of generality.
Then the Hubble parameter in the $x$-direction, $H_{x}$, must diverge to positive infinity in the limit $t \to 0$ (see the statement around Eq.\,\eqref{eqK-84}).
At the same time, $H_{x}$ should satisfy Eq.\,\eqref{eqI-24} to make the past boundary extendible.
As expressed in Eq.\,\eqref{eqA-7}, the leading behavior of such solutions around $t = 0$ can be summarized as 
\begin{align}
  H_{x}(t) = \frac{1}{t} + \mathcal{O}(t^{0}) \,.
  \label{eqI-31-3}
\end{align} 
Then the leading behavior of the scale factor in the $x$-direction is determined as $a_{x}(t) = t + \mathcal{O}(t^{2})$.
Taking account of this fact, let us write the scale factor in the $x$-direction around $t = 0$ as
\begin{align}
a_{x}(t) = t + \sum_{p=2}^{\infty} \frac{a^{(p)}_{x}}{p!} t^{p}
\label{eqI-31-4}
\end{align} 
where $a^{(p)}_{x}$ are coefficients independent of $t$.
Then imposing the condition \eqref{eqI-21} for $j = x$ leads to
\begin{align}
a^{(2)}_{x} = 0 \,.
\label{eqI-31-5}
\end{align} 
Next, from the condition \eqref{eqI-22}, the products $H_{x}H_{y}$ and $H_{x}H_{z}$ must be bounded in the limit $t \to 0$.
Combining this fact with Eq.\,\eqref{eqI-31-3}, we can see that $H_{y}$ and $H_{z}$ must vanish in the limit $t \to 0$ as fast as or faster than $t$, i.e.,
\begin{align}
H_{y,z} = \mathcal{O}(t)\,.
\label{eqI-34}
\end{align}   
Then the conditions \eqref{eqI-21} and \eqref{eqI-22} are all satisfied.
From Eq.\,\eqref{eqI-34}, now we can take the scale factors $a_{y,z}$ as the power series around $t=0$ as
\begin{align}
a_{y,z}(t) &= 1 + \sum_{p=2}^{\infty} \frac{a^{(p)}_{y,z}}{p!} t^{p} \,,
\label{eqI-34-3}
\end{align} 
where $a_{y,z}^{(p)}$ are coefficients independent of $t$.
Then the remaining condition \eqref{eqI-23} gives constraints as
\begin{align}
a_{y,z}^{(3)} = 0 \,.
\label{eqI-34-4}
\end{align} 

In summary, we can conclude that the past boundary is at least locally extendible if
\begin{align}
a_{x}(t) &= t + \mathcal{O}(t^{3}) \,,
\label{eqI-34-5}\\
a_{y,z}(t) &= 1 + \frac{a_{y,z}^{(2)}}{2} t^{2} + \mathcal{O}(t^{4}) \,,
\label{eqI-34-6}
\end{align} 
or equivalently,
\begin{align}
H_{x}(t) &= \frac{1}{t} + \mathcal{O}(t) \,, 
\label{eqI-34-7}\\
H_{y,z}(t) &= a_{y,z}^{(2)} t + \mathcal{O}(t^{3})\,.
\label{eqI-34-8}
\end{align}
Consequently, in the extendible case, the geometry is asymptotically given by the product of the two-dimensional Milne universe and $\mathbb{R}^2$ toward the boundary:
\begin{align}
  ds^2 \simeq -dt^2 + t^2 \, dx^2 + dy^2 + dz^2
  \quad (t \to 0)\,.
  \label{eqI-35}
\end{align}
 
As already mentioned, if a scale factor in a spatial direction vanishes at a finite time, the Hubble parameter in that direction must diverge as approaching the time.
Note that the divergence of the Hubble parameters in more than one spatial direction conflicts with the extendibility condition \eqref{eqI-22}.
Therefore, there inevitably exists a p.p.~curvature singularity on the past boundary if the scale factors in more than one spatial direction simultaneously vanish at a finite time.
This result is consistent with that of Subsec.\,\ref{subsec:Iso-3}, where it is found that FLRW spacetime whose scale factor vanishes at a finite time may be extendible only if it is spatially open, because the spatially open FLRW spacetime is included in Bianchi type V/$\text{VII}_{h}$, not the type I we are considering here.

The condition just obtained, Eq.\,\eqref{eqI-34-6}, forbids that the scale factors $a_{y,z}$ contain a term proportional to $t^3$ around $t=0$ for the extendibility.
To understand this, as a toy model, let us consider the following metric,
\begin{align}
  ds^{2}
  &= -\rd t^{2} + t^{2} \rd x^{2} + e^{\beta_{y} t^{3}} \rd y^{2} + e^{\beta_{z} t^{3}} \rd z^{2} \,,
  \label{eqI-40}
\end{align}
with constants $\beta_{y,z}$.
In this spacetime, comoving and null geodesics reach a past boundary at $t=0$.
The scale factor in the $t$-$x$ plane is linear with respect to time as in the Milne universe, and satisfies the condition \eqref{eqI-34-5}.
So, as performed in Eq.\,\eqref{eqK-75}, one would try to introduce new coordinates $\{T, X\}$ as
\begin{align}
T = t \cosh x \,, \quad 
X = t \sinh x \,.
\label{eqI-45}
\end{align} 
Then the metric \eqref{eqI-40} is rewritten as
\begin{align}
ds^{2} = - \rd T^{2} + \rd X^{2} + e^{\beta_{y}(T^{2} - X^{2})^{3/2}} \rd y^{2} + e^{\beta_{z}(T^{2} - X^{2})^{3/2}} \rd z^{2} \,.
\label{eqI-46}
\end{align}
As expected, the $T$-$X$ plane is completely well-behaved.
However, not all of the metric components are differentiable on the original past boundary, which is now located at $T=X$, due to the factor of $(T^{2} - X^{2})^{3/2}$ in the exponential.
Thus the spacetime is not extendible beyond the boundary.

\subsubsection{Comoving complete but null incomplete case: $t_{i} = - \infty$ and $\lambda(t_{i}) = \text{finite}$}
\label{subsec:Aniso-3-tinf}

In a case of $t_{i} = - \infty$ while $\lambda(t_{i}) = \text{finite}$, where $\lambda(t_i)$ is the affine parameter of a null geodesic in the $x$-direction at $t_i=-\infty$ (see Eq.\,\eqref{eqI-10-2}), first we need to find $H_{j}$ with $j = x,y,z$ satisfying Eq.\,\eqref{eqI-21} in the limit $t \to - \infty$.
As discussed in Appendix \ref{appA}, such a function should converge to some constant in the limit:
\begin{align}
\lim_{t \to -\infty} H_{j}(t) = \bar{H}_{j}
\label{eqI-61}
\end{align} 
where $\bar{H}_{j}~(j=x,y,z)$ are non-negative constants.
Then the condition \eqref{eqI-22} is automatically satisfied.
Note that from our assumption, we need
\begin{align}
\bar{H}_{x} \geq 0 \,.
\label{eqI-62}
\end{align}
Now the additional condition \eqref{eqI-23} implies
\begin{align}
\bar{H}_{y,z}(\bar{H}_{y,z} - \bar{H}_{x}) = 0\,,
\label{eqI-62-2}
\end{align} 
from which we can see that the past boundary is extendible only in the following cases:
\begin{enumerate}[(a)]
\item
$\bar{H}_{x} = \bar{H}_{y} = \bar{H}_{z}$; the universe is expanding spatially homogeneously and isotropically in the limit $t \to - \infty$.
When $\bar{H}_x$ is non-vanishing, the geometry approaches the four-dimensional flat de Sitter universe, $dS^4$.
\item
$\bar{H}_{x} = \bar{H}_{z} > 0$ and $ \bar{H}_{y} = 0$, or, $\bar{H}_{x} = \bar{H}_{y} > 0$ and $\bar{H}_{z} = 0$;
the universe expands in two directions at the same rate.
The scale factor in the remaining one direction does not evolve.
That is, the geometry is asymptotically $dS^3 \times \mathbb{R}$.
\item
$\bar{H}_{x}> 0$ and $\bar{H}_{y} = \bar{H}_{z} = 0$; the universe is expanding only in the $x$-direction. 
That is, the geometry is asymptotically $dS^2 \times \mathbb{R}^2$.
\end{enumerate} 

Note that if all the scale factors in three spatial directions have time-dependence in the limit $t \to - \infty$, the spacetime is extendible beyond the boundary only in the case (a), in which the spacetime asymptotically becomes nothing but the flat de Sitter universe.

Also, we should note that it is necessary, not sufficient, for local extendibility that the situation is included in any of the above cases (a)--(c).
In particular, the subleading behavior of the Hubble parameters (or scale factors) with respect to time can affect the extendibility.
Below, we consider a case that the deviation of the Hubble parameters from the constants is given by the negative power of $-t$ , and a case that the deviation is given by the exponential of $-t$.

\paragraph{Deviation by $(-t)^{-p}$.}
As a first example, let us consider a case that the Hubble parameters in the $x$ and $y$-directions converge to $\bar{H}_x$ and $\bar{H}_y$ as $t \to -\infty$, respectively, but both of them include the deviation proportional to the negative power of $-t$:
\begin{align}
H_{x}(t) &\simeq \bar{H}_x + \frac{b_{x}}{(-t)^{p_{x}}}\quad (p_{x}>0)\,,
\label{eqI-71} \\
H_{y}(t) &\simeq \bar{H}_y + \frac{b_{y}}{(-t)^{p_{y}}}\quad (p_{y}>0)\,,
\label{eqI-71.2}
\end{align}
where $b_{x}$ and $b_{y}$ are some constants.
Here, $\bar{H}_x$ and $\bar{H}_y$ are included in the above cases (a)--(c) as the necessity for the extendibility.
(The following discussion can be applied in the same way for the Hubble parameter in the $z$-direction.)

Given the Hubble parameters as Eqs.\,\eqref{eqI-71} and \eqref{eqI-71.2}, the conditions \eqref{eqI-21} and \eqref{eqI-22} can be satisfied under the appropriate choice of $H_z$.
However, the satisfaction of the condition \eqref{eqI-23} is nontrivial.
Corresponding to the above Hubble parameter, the scale factor in the $x$-direction is approximately given by
\begin{align}
a_{x}(t) &\simeq \begin{dcases}
  e^{\bar{H}_x t} \left(1 + \frac{b_x}{p_x - 1} \frac{1}{(-t)^{p_x -1}}\right) 
  &(p_x > 1) \,,\\
  e^{\bar{H}_x t} (-t)^{-b_{x}}
  &(p_{x}=1)\,,\\
  e^{\bar{H}_x t} \exp\left( \frac{-b_{x}}{1-p_{x}} (-t)^{1-p_{x}} \right)
  &(0 < p_{x} < 1)\,,\\
\end{dcases} 
\label{eqI-72}
\end{align} 
where irrelevant terms as $t \to -\infty$ are neglected.

First, we consider the case of $\bar{H}_x > 0$, in which there is always a past boundary in the $x$-direction.
Focusing only on the dominant part as $t \to -\infty$, the quantity concerning the condition \eqref{eqI-23} is approximately given by
\begin{align}
\frac{1}{a_{x}^{2}}\left( \dot{H}_{y} + H_{y}^{2} - H_{x}H_{y} \right)
&\sim
  e^{-2\bar{H}_x t} \bigg[ \bar{H}_y (\bar{H}_y - \bar{H}_x) - \frac{\bar{H}_y b_x}{(-t)^{p_x}} + \frac{(2\bar{H}_y - \bar{H}_x) b_y }{(-t)^{p_y}} \notag\\
&\qquad \qquad+ \frac{b_y^2}{(-t)^{2p_y}} - \frac{b_xb_y}{(-t)^{p_x +p_y}} + \frac{b_yp_y}{(-t)^{1+p_y}} \bigg] \quad (t \to -\infty) \,,
\label{eqI-73}
\end{align} 
where the minor time dependence irrelevant to our analysis is suppressed.
From this expression, we can obtain the following results depending on the relation between $\bar{H}_x$ and $\bar{H}_y$:

\begin{enumerate}[(i)]
  \item $\bar{H}_x = \bar{H}_y > 0$

  This case is included in the above case (a) or (b) with appropriate $H_z$.
  But, in this case, the quantity \eqref{eqI-73} is necessarily diverges as $t \to -\infty$ unless $b_x = 0$ and $b_y = 0$.
  Thus, the past boundary located at $t = - \infty$ in the $x$-direction is the p.p.~curvature singularity unless $b_x = 0$ and $b_y = 0$.

  \item $\bar{H}_x > 0$ and $\bar{H}_y = 0$
  
  This case is included in the above case (b) or (c) with appropriate $H_z$.
  But, in this case, the quantity \eqref{eqI-73} is necessarily diverges unless $b_y = 0$.
  Thus, the past boundary located at $t = - \infty$ in the $x$-direction is the p.p.~curvature singularity unless $b_y = 0$.
  With $b_y = 0$, i.e. $a_y =1$, the nontrivial deviation of $a_x$ as Eq.\,\eqref{eqI-72} is compatible with the extendibility, which corresponds to 1) in Table \ref{tab:extendibleBianchiI}.
\end{enumerate}

Next, let us consider the case of $\bar{H}_x = \bar{H}_y = 0$, which fits into the above case (a) with appropriate $H_z$.
In this case, we require $b_{x} > 0$ in Eq.\,\eqref{eqI-71} since we are interested in the universe expanding in the $x$-direction.
When $p_{x}>1$, $a_{x}$ does not vanish as $t \to -\infty$, which is out of our interest.
Thus we consider the situation with $b_x > 0$ and $0 < p_x \leq 1$.
Then, we have
\begin{align}
&\frac{1}{a_{x}^{2}} \left( \dot{H}_{y} + H_{y}^{2} - H_{x}H_{y} \right)
\notag \\
&\simeq
\begin{dcases}
  b_{y} \exp\left( \frac{2b_{x}}{1-p_{x}}(-t)^{1-p_{x}} \right)\left[ \frac{p_{y}}{(-t)^{1+p_{y}}} -\frac{b_{x}}{(-t)^{p_{x}+p_{y}}} + \frac{b_{y}}{(-t)^{2p_{y}}} \right]
  &(0 < p_x < 1)\,,\\
  b_y \left[ (p_y-b_x) (-t)^{2b_x - p_y -1} + b_y (-t)^{2b_x - 2p_y} \right]
  &(p_x = 1)\,,\\
\end{dcases}
\label{eqI-77}
\end{align} 
from which we get the following result:

\begin{enumerate}[(i)]
  \setcounter{enumi}{2}
  \item  $\bar{H}_x = \bar{H}_y = 0$

  When $p_{x} = 1$, it suffices to consider the case with $b_{x} > 1$, otherwise the spacetime is geodesically complete in the past null (and also comoving) direction in the $t$-$x$ plane (as in FLRW spacetime discussed in Subsec.\,\ref{subsec:Iso-3}) which is out of our interest.
  In the case with $p_x = 1$ and $b_x > 1$, the condition \eqref{eqI-23} can be satisfied only if
  \begin{align}
  \begin{dcases}
  b_{y} = 0 &\text{if} ~ 0 < p_{y} < 1\,, \\
  b_{y} = 0\quad \text{or} \quad b_{x}-b_{y}=1 &\text{if} ~ p_{y} =1\,, \\
  b_{y}=0 \quad \text{or} \quad b_{x}=p_{y} \quad \text{or} \quad p_{y} \geq 2b_x -1 &\text{if} ~ p_{y} >1\,.
  \end{dcases}
  \label{eqI-78}
  \end{align} 
  From this result, for example, we can consider scale factors which have nontrivial time-dependence and lead to an extendible past boundary at $t = - \infty$ for null geodesics in the $x$-direction as follows:
  \begin{align}
    a_x(t) \simeq \frac{1}{(-t)^{b_x}} \,, \quad
    a_{y,z}(t) \simeq 
    \begin{dcases}
      \frac{1}{(-t)^{b_x -1}} \,,\\
      1 + \frac{b_{y,z}}{p_{y,z} -1} \frac{1}{(-t)^{p_{y,z} -1 }} \quad (b_x = p_{y,z} ~ \text{or} ~ p_{y,z} \geq 2b_x -1)\,,
    \end{dcases}
    \label{eqI-78.2}
  \end{align}
  where $b_x > 1$.
  For the extension of spacetime with these scale factors, see Appendix \ref{appB}.
  In these specific cases, note that the time-dependence in higher order beyond Eq.\,\eqref{eqI-78.2} may affect the extendibility.
  Extendible cases with the nontrivial deviation included in Eq.\,\eqref{eqI-78} are given by 2), 3), and 4) in Table \ref{tab:extendibleBianchiI}.
  Notice that null geodesics toward the $y$-direction are incomplete in the case 3) with $b_x > 2$.
  Then, since the Hubble parameters do not satisfy the extendibility condition \eqref{eqI-23} with swapping $x$ and $y$, this boundary is a p.p.~curvature singularity.

  When $0 < p_{x} < 1$, Eq.\,\eqref{eqI-77} diverges as $t \to -\infty$ and thus the condition \eqref{eqI-23} is violated unless $b_{y} = 0$.
  With $b_y = 0$, i.e. $a_y =1$, any deviation of $H_x$ from zero is compatible with the extendibility, which corresponds to 5) in Table \ref{tab:extendibleBianchiI}.

\end{enumerate}

In summary, if there are some deviation of the Hubble parameters from the constants by the negative power of $-t$, a non-scalar curvature singularity at $a_{x} = 0$ is induced in general due to the violation of the condition \eqref{eqI-23}, and extendible cases are highly restricted as studied above. 

\begin{table}[t]
  \centering
  \begin{tabular}{l|l}
  \multicolumn{2}{l}{\textbf{Extendible Bianchi I}}                                                            \\   \hline
                                              & scale factor                                                             \\ \hline
  comoving and null incomplete:               & $a_x=t + \mathcal{O}(t^3)$                                            \\ 
  $a_x(0) = 0$                              & $a_{y,z}=1 + \alpha_{y,z} t^2 + \mathcal{O}(t^4)$                            \\ \hline
  null incomplete: $a_x(-\infty) =0$        &1) $a_x = e^{\bar{H}_xt + \cdots} , ~a_{y,z} = 1$ \\
  and $\int^{-\infty}a_x\,dt = \text{finite}$ &2) $a_x = (-t)^{-b_x} + \cdots , ~a_{y,z} = 1,$ with $b_x > 1$\\
  &3) $a_x = (-t)^{-b_x}, ~a_{y,z} = (-t)^{-b_x +1 },$ with $1 < b_x \leq 2$\\ 
  &4) $a_x = (-t)^{-b_x}, ~a_{y,z} = 1 + \alpha_{y,z} (-t)^{- p_{y,z}+1} $,\\
  &\quad with $b_x > 1$ and ($b_x = p_{y,z}$ or $p_{y,z} \geq 2 b_x -1 $)\\
  &5) $a_x = \exp(\beta_x (-t)^{1-p_x} + \cdots), ~ a_{y,z} = 1$,\\
  &\quad  with $\beta_x < 0$ and $0< p_x < 1$\\
  &6) $a_x = e^{\bar{H} t} + \alpha_{x}e^{(\bar{H} + c_x) t} + \cdots $,\\
  &\quad $a_{y,z} = e^{\bar{H} t} + \alpha_{y,z}e^{(\bar{H} + c_{y,z}) t} + \cdots $, with $\mathrm{min}\{ c_x, c_{y,z} \} \geq 2\bar{H}$\\
  &7) $a_x = e^{\bar{H}_x t} + \alpha_{x}e^{(\bar{H}_x + c_x) t} + \cdots $,\\
  &\quad $a_{y,z} = 1 + \alpha_{y,z}e^{c_{y,z} t} + \cdots $\,, with $c_{y,z} \geq 2\bar{H}_x$\\\hline
  \end{tabular}
  \caption{Bianchi type I spacetimes with at least locally extendible past boundary (an apparent coordinate singularity) at $a_x = 0$. Here, $\alpha_{x,y,z}$ are arbitrary constants.}
  \label{tab:extendibleBianchiI}
\end{table}

\paragraph{Deviation by $e^{ct}~(c>0)$.}

Let us consider another example such that the Hubble parameters in the $x$ and $y$-directions are given by the constants with the deviation suppressed exponentially as $t \to -\infty$:
\begin{align}
H_{x}(t) &\simeq \bar{H}_x + d_{x} e^{c_{x}t} \quad (c_x > 0)\,,
\label{eqI-81} \\
H_{y}(t) &\simeq \bar{H}_y + d_{y} e^{c_{y}t} \quad (c_y > 0)\,,
\label{eqI-81.2}
\end{align} 
where $d_x$ and $d_y$ are constants.
(Again, the same argument holds for the Hubble parameter in the $z$-direction.)
Here we take $\bar{H}_x$ to be a positive constant.
Then the conditions \eqref{eqI-21} and \eqref{eqI-22} are satisfied.
On the other hand, the scale factor in the $x$-direction is appropriately given by
\begin{align}
a_{x}(t) &\simeq e^{\bar{H}_x t} + \frac{d_x}{c_x} e^{(\bar{H}_x + c_x) t} \,,
\label{eqI-82}
\end{align} 
so we have
\begin{align}
&\frac{1}{a_{x}^{2}} (\dot{H}_{y} + H_{y}^{2} - H_{x}H_{y})
\notag \\
&\simeq  \bar{H}_y (\bar{H}_y - \bar{H}_x ) e^{-2\bar{H}_x t} + d_y (2\bar{H}_y - \bar{H}_x + c_y) e^{(c_y -2\bar{H}_x )t} - \bar{H}_y d_x e^{(c_x-2\bar{H}_x) t} + \cdots
\,,
\label{eqI-83}
\end{align} 
where the terms which may be dominant as $t \to -\infty$ are extracted.
To keep the past boundary extendible, this quantity must be bounded as $t \to -\infty$.
Note that the first term in Eq.\,\eqref{eqI-83} vanishes since we consider any of the above cases (a)--(c).
Thus the requirements for the extendibility are summarized as follows depending on the relation between $\bar{H}_x$ and $\bar{H}_y$:

\begin{enumerate}[(i)]
  \item $\bar{H}_x = \bar{H}_y > 0$
  
  This case is included in the above case (a) or (b) with appropriate $H_z$.
  From Eq.\,\eqref{eqI-83}, the local extendibility still holds if
  \begin{align}
    \mathrm{min}\{ c_x, c_y \} \geq 2\bar{H}_x\,,
    \label{eqI-83.2}
  \end{align}
  which is the case 6) in Table \ref{tab:extendibleBianchiI}.
  If not, the coefficient of $e^{(\mathrm{min}\{ c_x, c_y \}-2\bar{H}_x ) t}$ must vanish, and we need to focus on the exponential of the next order.

  \item $\bar{H}_x > 0$ and $\bar{H}_y = 0$
  
  This case is included in the above case (b) or (c) with appropriate $H_z$.
  The local extendibility is ensured if
  \begin{align}
    c_y \geq 2\bar{H}_x\,,
    \label{eqI-83.3}
  \end{align}
  which is the case 7) in Table \ref{tab:extendibleBianchiI}.
  If not, the coefficient of $e^{(c_y - 2\bar{H}_x ) t}$, i.e. $d_y (c_y - \bar{H}_x) $ must vanish, and we need to focus on the exponential of the next order.
\end{enumerate}

\section{Summary and discussion}
\label{sec:Summary}

In this paper, we considered cosmological spacetimes with a past boundary, where the scale factor (in at least one spatial direction) vanishes and past-directed causal geodesics reach at a finite affine parameter.
We investigated when the parallelly propagated (p.p.) curvature singularity occurs by checking the behavior of the curvature tensor in a p.p.~tetrad basis along the geodesic. 
This paper was devoted to the analysis for Friedmann--Lema\^{i}tre--Robertson--Walker (FLRW) spacetime with a spatial curvature, and for Bianchi type I spacetime.

In Sec.\,\ref{sec:Isotropic}, we considered FLRW spacetime with a spatial curvature and studied p.p.~curvature singularity or extendibility of a past boundary for causal geodesics.
Then we found that the past boundary is extendible at least locally if the conditions \eqref{eqK-81} and \eqref{eqK-82} are satisfied.
Table \ref{table1} gives the classification of the singularity on the past boundary.
We clarified that FLRW spacetime incomplete in both comoving and null directions can be extendible at least locally only if it asymptotically reduces to the Milne universe as going back to the past, while FLRW spacetime incomplete in the null direction (while complete in the comoving direction) can be extendible at least locally only if the metric asymptotically takes the flat de Sitter's form as $t \to -\infty$.
Furthermore, the extendibility is sensitive to the subleading time-dependence of the scale factor as summarized in Table \ref{tab:extendibleFLRW}.

In Sec.\,\ref{sec:Anisotropic}, as the simplest extension to spacetime with anisotropy, we studied p.p.~curvature singularity or extendibility of a past boundary in Bianchi type I spacetime.
The local extendibility conditions of the past boundary are given by Eqs.\,\eqref{eqI-17}--\eqref{eqI-20}, and the classification of the singularity is summarized as Table \ref{table2}. 
At least locally extendible cases are listed in Table \ref{tab:extendibleBianchiI}.
In particular, when the spacetime is comoving and null incomplete, the boundary is extendible only if the scale factors are asymptotically given by Eqs.\,\eqref{eqI-34-5} and \eqref{eqI-34-6} as going back to the past.
When the spacetime is comoving complete but null incomplete, the Hubble parameter in each direction must be constant and the expanding dimensions must be isotropic toward the past to avoid the singularity.
In other words, the absence of the singularity requires highly symmetric geometry of spacetime, such as $dS^{4}$, $dS^{3} \times \mathbb{R}$, and $dS^{2} \times \mathbb{R}^{2}$.
When the Hubble parameters in all directions vanish as $t \to -\infty$, some specific cases such as Eq.\,\eqref{eqI-78.2} have an extendible past boundary in the null direction.

As a future direction, it is important to investigate the p.p.~curvature singularity or extendibility in more general structure of spacetime.
First, as the direct extension of the present work, we should reveal singularity in general anisotropic cosmologies.
In particular, Bianchi V, VII$_{h}$, and IX spacetimes include spatially open and closed FLRW spacetimes as special cases, so it will be intriguing to search extendibility conditions of the past boundary in such spacetimes. 
Another interesting structure of spacetime is periodicity in space.
For example, if one tries to extend the flat de Sitter universe with torus identification toward the null direction, the resulting universe exhibits some ill-behaviors, such as a closed timelike curve and no global extension\,\cite{Numasawa:2019juw}.
So it will be interesting to reveal similar phenomena in the anisotropic universe.
Furthermore, it may be useful to analyze the presence of the boundary and its extendibility in higher dimensions since the geometry of extra dimensions may be constrained by requiring the absence of singularity.

By focusing on the p.p.~curvature singularity, in this paper, we pointed out the possibility of non-scalar curvature singularity, where any scalar curvature invariant is bounded.
A comprehensive classification of the causal structure of FLRW spacetime based on the scalar curvature singularity can be seen in Ref.\,\cite{Harada:2018ikn}, but incorporating our investigation about the non-scalar curvature singularity will alter the picture.
The exhaustive analysis in more general geometry taking account of singularities other than the scalar curvature singularity will be laboring, but worthwhile.

Finally, we would like to comment that our results would provide a new guideline for formulating theories beyond GR which resolve the singularity.
Classical gravitational theories with limited curvature are expected to be effective theories of some fundamental theories such as quantum gravity. 
In Ref.\,\cite{Sakakihara:2020rdy}, a Bianchi type I solution without scalar curvature singularity is proposed in such a context, so-called \emph{limiting curvature theory}.
However, according to the criteria given in the present paper, it turns out that the proposed model has a non-scalar curvature singularity since the Hubble parameter in one direction is different from the others.
Therefore, a new mechanism will be needed to avoid any p.p.~curvature singularity.
In the context of loop quantum gravity, the resolution of singularity in cosmology has been actively studied \,\cite{Ashtekar:2006rx, Bojowald:2006da, Ashtekar:2009vc, Singh:2011gp}.
As we clarified the possibilities of the p.p.~curvature singularity which should be resolved, we hope that our analysis will be useful in constructing such theories beyond GR.

\acknowledgments
We thank Yuki Sakakihara and Tomohiro Harada for fruitful discussions.
K.N.~was supported by JSPS KAKENHI Grant Number JP21J20600.
D.Y.~was supported by JSPS KAKENHI Grant Numbers JP19J00294 and JP20K14469.

\appendix

\section{Solutions of $dH(t) / dt + H^{2}(t) = \mathrm{const.}$}
\label{appA}

In this Appendix, we study a differential equation appearing in Eqs.\,\eqref{eqK-83} and \eqref{eqI-24}, 
\begin{align}
\frac{dH(t)}{dt} = - H^{2}(t) + Q
\label{eqA-1}
\end{align} 
with a constant $Q$.
In particular, we classify the asymptotic behavior of the solutions when $t$ decreases since in this paper we are interested in how $H(t)$, which denotes the Hubble parameter, should behave as going back to the past to avoid the spacetime singularity.
As we will see below, the behavior of the solutions depends on whether $Q$ is positive, zero, or negative.
Let us investigate each case separately.

\begin{enumerate}[(i)]

\item
$Q > 0$

In this case, Eq.\,\eqref{eqA-1} can be factorized as $dH(t)/dt = -(H(t) + \sqrt{Q})(H(t) - \sqrt{Q})$.
It is obvious that it has trivial (constant) solutions, $H(t) = \pm \sqrt{Q}$.
Otherwise, from the factorization, one can see that there are three distinct flows of the solutions, see the first part of Fig.\,\ref{figA-1}.
One of the flows always with $dH(t)/dt > 0$ corresponds to the solution given by
\begin{align}
H(t) = \sqrt{Q} \tanh (\sqrt{Q} ( t -t_{1}))
\label{eqA-4}
\end{align} 
with a finite constant $t_{1}$, in which $t$ is defined within $(- \infty , \infty)$.
$H(t)$ is then bounded within $-\sqrt{Q} < H(t) < \sqrt{Q}$, and asymptotically goes to $\sqrt{Q}$ as $t \to \infty$, while goes to $-\sqrt{Q}$ as $t \to -\infty$.
Also there are flows with $dH(t)/dt < 0$, which are specified by the solution
\begin{align}
H(t) = \sqrt{Q} \coth (\sqrt{Q} (t - t_{1}') )
\label{eqA-5}	
\end{align} 
with a finite constant $t_{1}'$.
When $t$ is defined within $( t_{1}', \infty)$, $H(t)$ monotonically increases as $t$ decreases, and eventually diverges to positive infinity in the right-sided limit $t \to t_{1}'$.
We can consider another flow with the same functional form as Eq.\,\eqref{eqA-5} but now $t$ is defined within $(- \infty, t_{1}')$.
Then $H(t)$ monotonically increases as $t$ decreases, and converges to $- \sqrt{Q}$ in the limit $t \to - \infty$.

\item
$Q = 0$

In this case, Eq.\,\eqref{eqA-1} has a trivial solution, $H(t) = 0$.
Otherwise, the solution needs to be always monotonically decreasing with $t$, i.e., $dH(t)/dt <0$.
The nontrivial solution is given by
\begin{align}
H(t) = \frac{1}{t - t_{2}} \,,
\label{eqA-3}
\end{align} 
where $t_{2}$ is a constant.
As shown in the second part of Fig.\,\ref{figA-1}, there are two flows of the solution depending on the domain within which $t$ is defined.
One is that $t$ is defined within $(t_{2}, \infty)$.
Then $H(t)$ increases as $t$ decreases, and eventually diverges in the right-sided limit $t \to t_{2}$.
The other is that $t$ is defined within $(-\infty, t_{2})$.
Then $H(t)$ converges to zero as $t$ decreases, in which $t$ runs to $- \infty$.

\item
$Q < 0$

In this case, the solution is given by
\begin{align}
H(t) = \sqrt{-Q} \cot ( \sqrt{-Q}(t - t_{3}) )
\label{eqA-2} 
\end{align} 
with a constant $t_{3}$.
As shown in the third part of Fig.\,\ref{figA-1}, the flow of the solution is unique: $H(t)$ increases as $t$ decreases. The solution $H(t)$ necessarily diverges in the right-sided limit $t \to t_{3}$ if we consider $t \in (t_{3}, t_{3} + \pi/\sqrt{-Q})$.

\begin{figure}[tb]
  \centering
  \includegraphics[width=15cm]{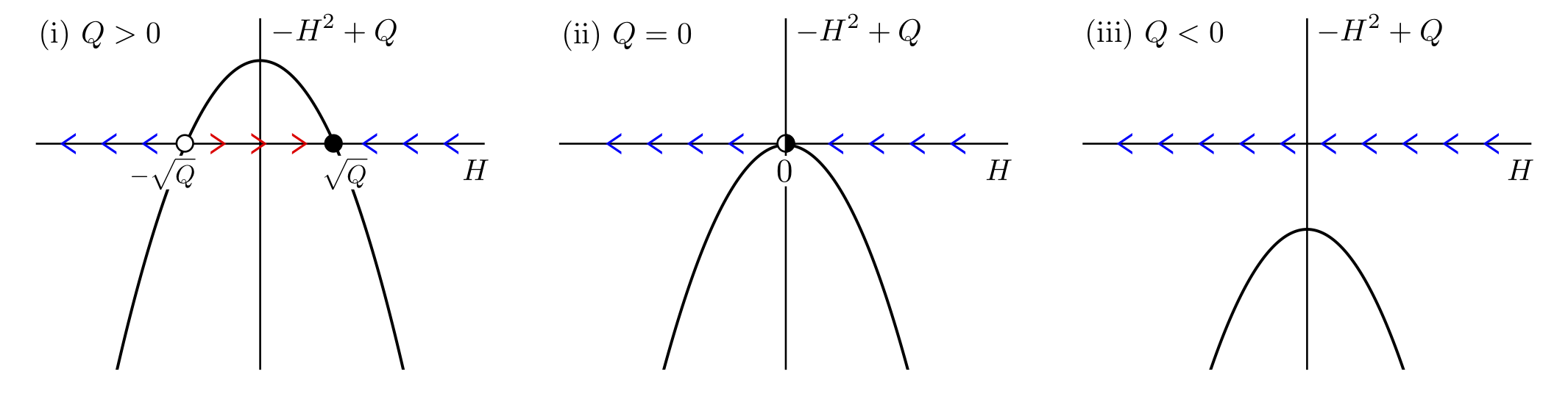}% Here is how to import EPS art
    \caption{Flows of the solutions of $dH(t)/dt = -H^{2}(t) + Q$. Figures (i), (ii), (iii) correspond to $Q >0, Q = 0, Q <0$, respectively. The arrows indicate the directions in which $H(t)$ flows when $t$ increases.}
  \label{figA-1}
\end{figure}
 
\end{enumerate} 

Let us collect consequences of the above analysis.
First, if we require that $H(t)$ diverges in the right-sided limit $t \to 0$ as $t$ decreases, $H(t)$ satisfying Eq.\,\eqref{eqA-1} is given by
\begin{align}
H(t) = 
\begin{dcases}
\sqrt{Q} \coth ( \sqrt{Q} \, t )
&(Q > 0)\,, \\
\frac{1}{t}
&(Q = 0)\,, \\
\sqrt{-Q} \cot ( \sqrt{-Q} \, t )
&(Q < 0)  \,.
\end{dcases} 
\label{eqA-6}
\end{align} 
Note that the solutions given by \eqref{eqA-6} can be expanded around $t=0$ as
\begin{align}
H(t) = \frac{1}{t} + \frac{Q}{3} t - \frac{Q^{2}}{45} t^{3} + \mathcal{O}(t^{5}) \,.
\label{eqA-7}
\end{align} 
Next, if we can define $t$ toward negative infinity keeping $H(t)$ continuous, then $H(t)$ satisfying Eq.\,\eqref{eqA-1} converges to some constant in the limit $t \to -\infty$.
Let us check this fact for cases (i), (ii), and (iii), one by one.
In the case (i), for which $H(t)$ should satisfy $dH(t)/dt = -H^{2}(t) + Q$ with positive $Q$, trivial solutions are given by $H(t) = \pm \sqrt{Q} = \text{const}$.
Nontrivial solutions \eqref{eqA-4} and \eqref{eqA-5} with $t$ running to $-\infty$ also exist, and they converge to $- \sqrt{Q}$ in the limit $t \to - \infty$.
In the case (ii), i.e. $Q = 0$, we have a trivial solution $H(t) = 0$ and a nontrivial one \eqref{eqA-3} with $t \in (-\infty ,t_{2})$.
However, the latter also approaches zero in the limit $t \to - \infty$.
In the case (iii), i.e. $Q < 0$, there is no solution such that $t$ can be defined to $- \infty$ keeping $H(t)$ continuous.
The analysis is finished.
We can conclude that the solutions of Eq.\,\eqref{eqA-1} necessarily converge to some constant in the limit $t \to -\infty$, if $t$ can be defined to $- \infty$.

\section{Past extension of extendible Bianchi I spacetime}
\label{appB}

In this Appendix, we clarify why the spacetimes classified to $2), 3)$ and $4)$ in Table \ref{tab:extendibleBianchiI} can be extended beyond the past boundary $t \rightarrow -\infty$ in the null direction.
For simplicity, we ignore the $z$-direction and consider just a three-dimensional spacetime here.

The Bianchi type I spacetime in the class $3)$ asymptotically reduces to the metric 
\begin{align}
g_{\mu\nu} dx^{\mu} dx^{\nu} = - dt^2 + \frac{1}{(-t)^{2b_{x}}} dx^2 + \frac{1}{(-t)^{2 b_{x} - 2}} dy^2 \qquad (b_{x} > 1)\label{eqB-1}
\end{align}
in the  limit $t \rightarrow - \infty$.
To perform the extension, we introduce coordinates $\lambda$ and $v$ by
\begin{align}
u & =  (b_{x} -  1) \int^{t} a_{x}(t) dt = \frac{1}{( - t)^{b_{x} -1}}\,, \\
v & = \frac{1}{b_{x} -  1} \left( x + \int^{t} \frac{dt}{a_{x}(t)} \right)\,.
\end{align}
Then, the metric components in $\{u,v,y\}$ coordinates are given by 
\begin{align}
 g_{\mu\nu} dx^{\mu} dx^{\nu} = - 2 du dv + (b_{x} - 1)^2 u^{\frac{2 b_{x}}{b_{x} - 1}} d v^2 + u^2 dy^2.
\end{align}
In the limit to the boundary, $u \rightarrow 0$, $g_{vv}$ can be ignored. Hence we obtain
\begin{align}
 g_{\mu\nu} dx^{\mu} dx^{\nu} \simeq - 2 du dv + u^2 dy^2 \qquad ( u \rightarrow 0)\,.
 \label{Mink_planewave}
\end{align}
Apparently, $g_{yy}$ looks singular in the limit $u \rightarrow 0$. However, this is just a coordinate singularity because Eq.\,\eqref{Mink_planewave} is the Minkowski metric in the Rosen's plane wave coordinates \cite{Blau:2002mw}. Actually, the coordinate transformation to the global Minkowski coordinates, $g_{\mu\nu} dx^{\mu} dx^{\nu} = - dT^2 + dX^2 + dY^2$, can be given by
\begin{align}
 T = \frac{1}{\sqrt{2}} \left( \left(1 + \frac{y^2}{2}\right) u + v \right), \qquad 
 X = \frac{1}{\sqrt{2}} \left( \left(1 - \frac{y^2}{2}\right) u - v \right) , \qquad 
 Y = u y.
\end{align}
Therefore the Bianchi type I spacetime \eqref{eqB-1} is extendible beyond the past boundary. 

A similar analysis holds for the spacetime in the class $2)$ and $4)$, where the metric asymptotically reduces to
\begin{align}
 g_{\mu\nu} dx^{\mu} dx^{\nu} = -dt^2 + \frac{1}{( - t )^{2 b_{x}}} dx^2 + dy^2 \,.
\end{align}
The metric components in the $\{u,v,y\}$ coordinates can be represented as
\begin{align}
 g_{\mu\nu} dx^{\mu} dx^{\nu} &= - 2 du dv + (b_{x} - 1)^2 u^{\frac{2 b_{x}}{b_{x} - 1}} d v^2 + dy^2 \notag\\
 &\simeq 
- 2 du dv + dy^2 \qquad ( u \rightarrow 0) \,.
\end{align}
The final expression is the Minkowski metric in the null coordinates and hence $ u = 0$ is just a coordinate singularity.

\bibliographystyle{JHEP}
\bibliography{biblio}% Produces the bibliography via BibTeX.

% Please avoid comments such as "For a review'', "For some examples",
% "and references therein" or move them in the text. In general,
% please leave only references in the bibliography and move all
% accessory text in footnotes.

% Also, please have only one work for each \bibitem.

\end{document}